\documentclass[twocolumn, longbibliography]{revtex4-2}

\newcommand*{\Title}{Exceptional Luttinger Liquids from sublattice dependent interaction}
\newcommand*{\Author}{Joachim Schwardt}

\usepackage{color}
\usepackage{graphicx}
\usepackage{gensymb}
\usepackage{amsfonts}
\usepackage{cancel}
\usepackage{amssymb}
\usepackage{amsmath}
\usepackage{amsthm}
\usepackage{array}
\usepackage{esint}
\usepackage{bm, bbm} 
\usepackage{booktabs}
\usepackage{multirow}
\usepackage{tikz}
\usepackage{pgffor}
\usetikzlibrary{calc, arrows.meta,arrows, graphs, quotes, positioning, angles, babel}
\usepackage{svg}
\usepackage{hyperref}
\usepackage{siunitx}
\usepackage{physics}
\usepackage{tensor}
\usepackage{slashed}
\usepackage[compat=1.1.0]{tikz-feynman}
\usepackage{csquotes}
\usepackage{mathtools}
\usepackage[all]{nowidow}

\makeatletter
\newcommand{\citecomment}[2][]{\citealp{#2}#1\citevar}
\newcommand{\citeone}[1]{\citecomment{#1}}
\newcommand{\citetwo}[2][]{\citecomment[,~#1]{#2}}
\newcommand{\citevar}{\@ifnextchar\bgroup{;~\citeone}{\@ifnextchar[{;~\citetwo}{]}}}
\newcommand{\citefirst}{\@ifnextchar\bgroup{\citeone}{\@ifnextchar[{\citetwo}{]}}}
\newcommand{\cites}{[\citefirst}
\makeatother

\let\abs\undefined
\let\bra\undefined
\DeclarePairedDelimiter\abs{\vert}{\vert}
\DeclarePairedDelimiter\kla{(}{)}
\DeclarePairedDelimiter\klb{[}{]}
\DeclarePairedDelimiter\klc{\{}{\}}
\DeclarePairedDelimiter\bra{\langle}{\rangle}
\DeclarePairedDelimiterX\brb[2]{\langle #1}{\rangle}{\delimsize\vert #2 }
\DeclarePairedDelimiterX\brc[3]{\langle #1}{\rangle}{\delimsize\vert #2 \delimsize\vert #3 }
\DeclareMathOperator{\sgn}{sgn}

\newcommand\numberthis{\addtocounter{equation}{1}\tag{\theequation}}

\newcommand{\intx}[4]{\int_{#1}^{#2} #3 \,\mathrm{d}#4}
\newcommand{\intr}[2]{\int_{\mathbb{R}} #1 \, \mathrm{d}#2}

\newcommand{\powe}[1]{\e^{#1}}

\DeclarePairedDelimiter\Ket{\vert}{\rangle}

\newcommand*{\F}{\mathcal{F}}

\newcommand{\e}{\text{e}}
\newcommand{\A}{\text{A}}
\newcommand{\B}{\text{B}}
\newcommand{\R}{\text{R}}
\newcommand{\M}{\text{M}}

\newcommand{\f}{\text{f}}
\newcommand{\K}{\text{K}}
\newcommand{\hc}{\text{h.c.}}

\newcommand{\vf}{v_\F}

\newcommand{\gray}[1]{\bgroup\color{white!40!black}#1\egroup}

\newcommand{\normord}[1]{
  {:\mathrel{\mspace{1mu}#1\mspace{1mu}}:}
}
\newcommand{\varnormord}[1]{
  {\mathrel{\mspace{1mu}\substack{\ast\\[-1.5pt] \ast}#1\substack{\ast\\[-1.5pt] \ast}\mspace{1mu}}}
}

\newcommand{\bsplit}[1]{\begin{aligned}[t]#1\end{aligned}}
\usepackage[english]{babel}
\usepackage[T1]{fontenc}
\usepackage{microtype}
\usepackage[english,capitalise]{cleveref}

\usepackage[acronym]{glossaries}
\makenoidxglossaries

\makeatletter
\let\oldtheequation\theequation
\renewcommand\tagform@[1]{\maketag@@@{\ignorespaces#1\unskip\@@italiccorr}}
\renewcommand\theequation{(\oldtheequation)}
\makeatother

\newcommand{\appref}[1]{\hyperref[#1]{Appendix~\ref*{#1}}}
\newcommand{\refcite}[1]{Ref.~\cite{#1}}

\renewcommand{\d}{\text{d}}
\renewcommand{\r}{\text{r}}

\renewcommand{\F}{\text{F}}

\renewcommand{\i}{\text{i}}
\renewcommand{\L}{\text{L}}
\renewcommand{\K}{\hat{\mathcal{K}}}
\renewcommand{\S}{\hat{\mathcal{S}}}

\newcommand{\U}{\hat{\mathcal{U}}}
\newcommand{\T}{\hat{\mathcal{T}}}
\newcommand{\C}{\hat{\mathcal{C}}}
\newcommand{\ret}{\text{ret}}
\newcommand{\PTR}{R}
\newcommand{\eplus}{^{\scriptscriptstyle{(+)}}}
\newcommand{\eminus}{^{\scriptscriptstyle{(-)}}}

\newcommand{\asymeq}[1]{\overset{#1}{\simeq}}
\newcommand{\asympropto}[1]{\overset{#1}{\sim}}
\newcommand{\tint}{\text{int}}

\newcommand{\spage}{p.\,}

\newcommand{\ssec}{sec.\,}

\newcommand{\sapp}{app.\,}
\newcommand{\seqn}{eqn.\,}

\newcommand{\sact}{\mathcal{S}}
\newcommand{\sgf}{G}

\newcommand{\thermit}{Hermitian}

\newcommand{\chs}{\hat{\mathcal{S}}}

\newcommand{\transpose}{{\mkern-1.5mu\text{T}}}
\newcommand{\eff}{\text{eff}}
\newcommand{\nameKallen}{Källén}

\makeatletter
\newcommand*{\glsplainhyperlink}[2]{%
    \begingroup%
      \hypersetup{hidelinks}%
      \hyperlink{#1}{#2}%
    \endgroup%
}
\let\@glslink\glsplainhyperlink
\makeatother
\hypersetup{
    colorlinks,
    breaklinks,
    citecolor=blue,
    linkcolor=blue,
    urlcolor=blue,
    pdfauthor={\Author},
    pdftitle={\Title}
}
\usetikzlibrary{decorations.pathmorphing}
\usetikzlibrary{decorations.markings}

\tikzfeynmanset{ 
arrow/.style={thick, shorten >=5pt,shorten <=5pt,->},
photon/.style={decorate, decoration={snake}, style={thick}},
electron/.style = {
decoration={
markings,
mark=at position 0.5 with {\arrow[xshift=1mm]{Stealth[width=1mm,length=2mm]}}},
postaction=decorate,
style={thick}},
electronNoMu/.style = {
decoration={
markings,
mark=at position 0.5 with {\arrow[xshift=1mm]{Stealth[width=1mm,length=2mm]}}},
postaction=decorate,
style = {thick, dashed} }
}

\newcommand{\SSoOneNoLabel}[1]{\begin{tikzpicture}[baseline={(0,0)}]
\begin{scope}[scale=#1]
\begin{feynman}
\fill (0,0) circle (10pt) coordinate (a);
\fill (5,0) circle (10pt) coordinate (b);
\fill (10,0) circle (10pt) coordinate (c);
\fill (15,0) circle (10pt) coordinate (d);
\diagram* {
(d) --[electron] (c),
(c) --[electron] (b),
(b) --[electron] (a),
(d) --[photon, half left] (b),
(c) -- [photon, half right] (a)
};
\end{feynman}
\end{scope}
\end{tikzpicture}}

\newcommand{\SSoTwoNoLabel}[1]{\begin{tikzpicture}[baseline={(0,0)}]
\begin{scope}[scale=#1]
\begin{feynman}
\fill (0,0) circle (10pt) coordinate (a);
\fill (0,5) circle (10pt) coordinate (b);
\fill (5,5) circle (10pt) coordinate (c);
\fill (5,0) circle (10pt) coordinate (d);
\diagram* {
(d) --[electron] (a),
(d) --[photon] (c) -- [electron, half right] (b) -- [electron, half right] (c),
(b) --[photon] (a)
};
\end{feynman}
\end{scope}
\end{tikzpicture}}

\newcommand{\GfreeU}[1]{\begin{tikzpicture}[baseline={(0,0)}]
\begin{scope}[scale=#1]
\begin{feynman}
\fill (0,0) circle (6pt) coordinate (a);
\fill (10,0) circle (6pt) coordinate (b);
\node [left] at (a) {\footnotesize $s$};
\node [right] at (b) {\footnotesize $s'$};
\diagram* {(b) --[electron, edge label= {\footnotesize $(k, \i \omega_n)$}] (a)};
\end{feynman}
\end{scope}
\end{tikzpicture}}

\newcommand{\Gfree}[1]{\begin{tikzpicture}[baseline={(0,0)}]
\begin{scope}[scale=#1]
\begin{feynman}
\fill (0,0) circle (6pt) coordinate (a);
\fill (10,0) circle (6pt) coordinate (b);
\node [left] at (a) {\footnotesize $s$};
\node [right] at (b) {\footnotesize $s'$};
\diagram* {(b) --[electronNoMu, edge label= {\footnotesize $(k, \i \omega_n)$}] (a)};
\end{feynman}
\end{scope}
\end{tikzpicture}}

\newcommand{\Vint}[1]{\begin{tikzpicture}[baseline={(0,0)}]
\begin{scope}[scale=#1]
\begin{feynman}
\fill (0,0) circle (6pt) coordinate (a);
\fill (10,0) circle (6pt) coordinate (b);
\node [left] at (a) {\footnotesize $s$};
\node [right] at (b) {\footnotesize $s'$};
\diagram* {(b) --[photon, edge label= {\footnotesize $(k, \i \omega_n)$}] (a)};
\end{feynman}
\end{scope}
\end{tikzpicture}}

\newcommand{\SHTnoLabel}[1]{\begin{tikzpicture}[baseline={(0,0)}]
\begin{scope}[scale=#1]
\begin{feynman}
\fill (0,0) circle (10pt) coordinate (a);
\fill (0,5) circle (10pt) coordinate (b);
\coordinate (c) at (0,8);
\diagram* {
(a) --[photon]  (b) -- [half right, out= -90, in=-90] (c) -- [electron, half right, out= -90, in=-90] (b)};
\end{feynman}
\end{scope}
\end{tikzpicture}}

\newcommand{\SFnoLabel}[1]{\begin{tikzpicture}[baseline={(0,0)}]
\begin{scope}[scale=#1]
\begin{feynman}
\fill (0,0) circle (10pt) coordinate (a);
\fill (5,0) circle (10pt) coordinate (b);
\diagram* {
(b) --[electron] (a) -- [photon, half left] (b)};
\end{feynman}
\end{scope}
\end{tikzpicture}}

\newcommand{\SSoOneLabelled}[1]{\begin{tikzpicture}[baseline={(0,0)}]
\begin{scope}[scale=#1]
\begin{feynman}
\fill (0,0) circle (4pt) coordinate (a);
\fill (3,0) circle (4pt) coordinate (b);
\fill (6.3,0) circle (4pt) coordinate (c);
\fill (9.2,0) circle (4pt) coordinate (d);
\node [left] at (a) {\footnotesize $s$};
\node [above] at (b) {\footnotesize $s'$};
\node [below] at (c) {\footnotesize $s$};
\node [right] at (d) {\footnotesize $s'$};
\diagram* {
(d) --[electronNoMu, near start, edge label'= {\tiny $(q_1, \i
 \nu_n)$}] (c),
(c) --[electronNoMu, edge label= {\tiny $(q_2, \i
 \nu_n')$}] (b),
(b) --[electronNoMu, near end, edge label= {\tiny $\substack{(k - q_1 + q_2, \\ \i \omega_n - \i
 \nu_n + \i
 \nu_n')}$}] (a),
(d) --[photon, half left, edge label'= {\tiny $(k-q_1)$}] (b),
(c) -- [photon, half right,  edge label=  {\tiny $(q_1 - q_2)$}] (a)
};
\end{feynman}
\end{scope}
\end{tikzpicture}} 

\newcommand{\SSoTwoLabelled}[1]{\begin{tikzpicture}[baseline={(0,0)}]
\begin{scope}[scale=#1]
\begin{feynman}
\fill (0,0) circle (4pt) coordinate (a);
\fill (0,3.5) circle (4pt) coordinate (b);
\fill (4,3.5) circle (4pt) coordinate (c);
\fill (4,0) circle (4pt) coordinate (d);
\node [left] at (a) {\footnotesize $s$};
\node [left] at (b) {\footnotesize $s$};
\node [right] at (c) {\footnotesize $s'$};
\node [right] at (d) {\footnotesize $s'$};
\diagram* {
(d) --[electronNoMu, edge label= {\tiny $(q_1, \i
 \nu_n)$}] (a),
(d) --[photon,  edge label'= {\tiny $(k-q_1)$}] (c) -- [electronNoMu, half right, edge label'= {\tiny $\substack{(k - q_1 + q_2, \\ \i \omega_n - \i
 \nu_n + \i
 \nu_n')}$}] (b) -- [electronNoMu, half right, edge label= {\tiny $(q_2, \i \nu_n')$}] (c),
(b) --[photon, near end, edge label= {\tiny $(k-q_1)$}] (a)
};
\end{feynman}
\end{scope}
\end{tikzpicture}}

\newacronym{ll}{LL}{Luttinger Liquid}
\newacronym{ep}{EP}{Exceptional Point}
\newacronym{qft}{QFT}{Quantum Field Theory}
\newacronym{rg}{RG}{Renormalization Group}
\newacronym{gf}{GF}{Green Function}
\newacronym{uv}{UV}{ultraviolet}
\newacronym{ir}{IR}{infrared}
\newacronym{nh}{NH}{non-Hermitian}
\newacronym{ope}{OPE}{Operator Product Expansion}
\newacronym{sg}{SG}{sine-Gordon}
\newacronym{pt}{PT}{Perturbation Theory}
\newacronym{trs}{TRS}{time-reversal symmetry}
\newacronym{phs}{PHS}{particle-hole symmetry}
\newacronym{cs}{CS}{chiral symmetry}
\newacronym{ssh}{SSH}{Su--Schrieffer--Heeger}
\newacronym{ell}{ELL}{Exceptional Luttinger Liquid}
\newacronym{ft}{FT}{Fourier Transform}
\newacronym{stdpar}{SP}{Standard Parameters: $\alpha = \frac{1}{\vf} = 2$, $a=1$ and $J=t_x=1$, $t_y=\frac{1}{2}$}
\newacronym{ct}{CT}{Convolution Theorem}
\newacronym{csba}{CSBA}{Conserving Second Born Approximation}

\newcommand{\jcb}[1]{\bgroup\color{orange} JCB: #1\egroup}
\definecolor{dgreen}{rgb}{0.1,0.5,0.1}
\definecolor{babyblue}{rgb}{0.54, 0.81, 0.94}
\newcommand{\carl}[1]{\bgroup\color{dgreen} CL: #1\egroup}
\newcommand{\benmi}[1]{\bgroup\color{babyblue!70!black} BM: #1\egroup}
\newcommand{\jsc}[1]{\bgroup\color{purple!75!blue} JS: #1\egroup}
\newcommand{\TODO}[1]{\bgroup\color{red!60!black}TODO: #1\egroup}

\begin{document}

\title{\Title}
\author{\Author$^{1,2}$}
\email{joachim.schwardt@tu-dresden.de}
\author{Benjamin Michen$^1$}
\author{Carl Lehmann$^1$}
\author{Jan Carl Budich$^{1,2}$}

\affiliation{$^1$Institute of Theoretical Physics, Technische Universit\"{a}t Dresden and W\"{u}rzburg-Dresden Cluster of Excellence ct.qmat, 01062 Dresden, Germany}
\affiliation{$^2$Max Planck Institute for the Physics of Complex Systems, N\"{o}thnitzer Str.~38, 01187 Dresden, Germany}

\date{\today}

\begin{abstract}
We demonstrate how \glspl{ep} naturally occur in the \gls{ll} theory describing the low-energy excitations of a microscopic lattice model with sublattice dependent electron-electron interaction.
Upon bosonization, this sublattice dependence directly translates to a non-standard \acrlong{sg}-type term responsible for the non-Hermitian matrix structure of the single-particle \gls{gf}.
As the structure in the lifetime of excitations does not commute with the underlying free Bloch Hamiltonian, non-Hermitian topological properties of the single-particle \gls{gf} emerge -- despite our Hermitian model Hamiltonian.
Both finite temperature and a non-trivial Luttinger parameter $K\neq 1$ are required for the formation of \glspl{ep}, and their topological stability in one spatial dimension is guaranteed by the chiral symmetry of our model.
In the presence of the aforementioned \acrlong{sg}-term, we resort to leading order \gls{pt} to compute the single-particle \gls{gf}.
All qualitative findings derived within \gls{ll} theory are corroborated by comparison to both numerical simulations within the conserving second Born approximation, and, for weak interactions and high temperatures, by fermionic plain \gls{pt}.
In certain parameter regimes, quantitative agreement can be reached by a suitable parameter choice in the effective bosonized model.
\glsresetall

\end{abstract}

\maketitle

\section{Introduction}\label{sec:intro}
\gls{ll} \footnote{Tomonaga-Luttinger Liquids \cite{Bosonization_DelftSchoeller_1998,QImpurity_review_Furusaki_2005,LL_EntanglementEntropy_Furukawa_2010,NHT_TLL_ExpVals_Yamamoto_2022,NH_LL_dissipative_BetheAnsatz_Exponents_2023,TLL_Kitaev_Gamma_Gohlke_2024} (or even Landau-Luttinger Liquids \cite{FermiLiquid_LL_review_Voit_1995}), originate from the Tomonaga-Luttinger model \cite{Tomonaga_1950,Luttinger_1963,Bosonization_Schulz_1995,LL_FiniteT_FiniteL_Properties_Mattson_1997,Bosonization_LL_LectureNotes_Schulz_1998,FermiLiquid_Breakdown_Castellani_1999,Bosonization_FiniteT_Bowen_2001,LL_TLL_review_Schonhammer_2003,Bosonization_Senechal_2004,RG_LL_Boundary_Grap_2009} and are often simply referred to as \acrlongpl{ll} \cite{Bosonization_LowE_1d_fluids_Haldane_1981,Bosonization_LL_LectureNotes_Schulz_1998,Bosonization_Rao_2001,Bosonization_FiniteT_Bowen_2001,Bosonization_Senechal_2004,RG_LL_Boundary_Grap_2009} in analogy to Landau's Fermi Liquid \cite{FermiLiquid_Theory_Landau_1956,FermiLiquid_Oscillations_Landau_1957,FermiLiquid_Landau_Luttinger_1962,FermiLiquid_LL_review_Voit_1995,Bosonization_Schulz_1995,FermiLiquid_review_Neilson_1996,LL_FiniteT_FiniteL_Properties_Mattson_1997,Bosonization_LL_LectureNotes_Schulz_1998,Bosonization_Rao_2001,FermiLiquid_Vignale_2022}.} theory provides an important toolbox for describing the low energy properties of one-dimensional (1D) correlated electrons as an effective bosonic field theory \cite{Bosonization_LowE_1d_fluids_Haldane_1981,Bosonization_Shankar_1994,FermiLiquid_LL_review_Voit_1995,Bosonization_Schulz_1995,Bosonization_DelftSchoeller_1998,Bosonization_Gogolin_1999,Bosonization_Rao_2001,Bosonization_FiniteT_Bowen_2001,LL_TLL_review_Schonhammer_2003,Bosonization_Miranda_2003,Giamarchi_2004,Bosonization_Senechal_2004}.
In higher spatial dimensions, Landau's Fermi Liquid theory tells us that the lifetime of quasi-particle excitations generically diverges with temperature as $1/T^2$ at the Fermi surface \cite{FermiLiquid_Oscillations_Landau_1957,Lifetime_QP_Qian_2005,FermiLiquid_Vignale_2022}, thus justifying an independent particle approximation.
By contrast, in 1D the lifetime $\sim 1/T$ \cite{LL_Electron_Dephasing_Gornyi_2005,LL_Electron_Lifetime_LeHur_2006} of single-particle type electronic excitations defies a free effective description, which may be remedied by bosonization in the framework of \gls{ll} theory in a wide range of settings \cite{Bosonization_AndersonLocal_Giamarchi_1988,QHallEdgeStates_LL_Wen_1990,Bosonization_TwoChainScaling_Nersesyan_1993,LL_Phases_2LLs_Schulz_1996,LL_FiniteT_FiniteL_Properties_Mattson_1997,LL_LongRangeElectricInteractions_Giamarchi_2000,LL_CompetingOrders_Wu_2003,LL_ImpuritySingleQuantumWires_Dolcini_2005,QImpurity_review_Furusaki_2005,Bosonization_SG_interior_Imambekov_2008,RG_LL_Boundary_Grap_2009,LL_EntanglementEntropy_Furukawa_2010,Helical_Rashba_SpinOrbit_QSH_Stroem_2010,Helical_Edge_PhononBackscat_Budich_2012,Helical_RashbaImp_RG_Crepin_2012,Helical_SyntheticQWire_Japaridze_2013,Helical_RashbaNanowires_LowElectronDensity_Schmidt_2016,QImpurity_LL_Rylands_2016,Bosonization_SSH_Giamarchi_2023,TLL_Kitaev_Gamma_Gohlke_2024}.

Realistic experimental circumstances to some extent deviate from the ideal zero temperature fixed point, hence putting finite lifetime effects back on the table in any of the aforementioned scenarios.
When scattering rates acquire a non-trivial matrix structure in some orbital or spin degree of freedom, imaginary parts of excitation spectra may be promoted to intriguing \gls{nh} topological properties \cite{NHT_QWalk_Transition_Rudner_2009,NHT_AnomalousEdgeState_Lee_2016,NHT_SSH_phases_Lieu_2018,NH_TopologicalPhases_Gong_2018,NHT_SymmetryProtectedNodalPhases_Budich_2019,NHT_EPs_QMBP_Luitz_2019,NHT_SymmetryProtectedExceptRings_Chiral_Yoshida_2019,NH_Topology_Sensors_Budich_2020,NHT_SkinEffect_Okuma_2020,NH_Topology_SkinEffect_Lin_2023,NH_SkinEffect_QuasiParticles_NNint_Micallo_2023} such as exceptional degeneracies \cite{NH_Exceptional_Berry_2004,EP_Physics_Heiss_2012,NHT_FiniteLifetime_QuasiParticles_Kozii_2017,NH_Physics_review_Ashida_2020,
ExceptTop_RevModPhys.93.015005,NH_Heff_SelfEnergy_1d_Hubbard_Rausch_2021,EP_DynamicInduced_Quenched_Lehmann_2021,NHT_ExceptPhases_DisorderQWire_Michen_2021,Emergent_NH_models_Eek_2024}.

In the context of \glspl{ll}, despite the peculiar role of lifetime effects in 1D, \gls{nh} physics has thus far only been considered for $\mathcal{PT}$-symmetric Hamiltonians \cite{NH_LL_PT_RG_QCritical_Ashida_2017,NH_LL_PT_Quench_Dora_2020,NH_LL_Quenched_Correlators_Dubey_2023}, or by phenomenologically adding external dissipation terms \cite{NH_LL_dissipative_Lindblad_EP_Nakagawa_2021,NHT_TLL_ExpVals_Yamamoto_2022,NH_LL_dissipative_BetheAnsatz_Exponents_2023}.
Both of these cases have in common that the initial model is already explicitly \gls{nh}.

\begin{figure}[htp!]
\centering
\includegraphics{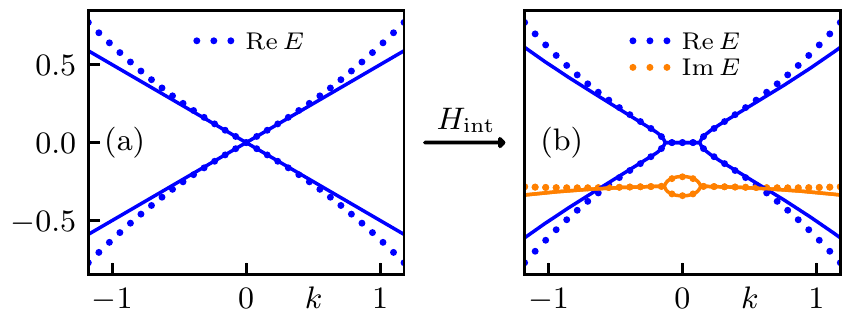}
\caption{Panel (a) shows the non-interacting dispersion of the free model \autoref{eqn:free_model}, with dots representing the exact result and the solid line corresponding to the linearization (with $\vf = t_ya = 0.5$).
As illustrated in panel (b), including sublattice dependent interactions \autoref{eqn:hint} leads to an effective \gls{nh} Hamiltonian and the formation of \glspl{ep} (here at $\omega=0$, $\beta=1$, $U_\A=1.5$ and $U_\B=1.1$).
Dots represent numerical data within \acrshort{csba}, and solid lines correspond to \gls{ll} \acrshort{pt} with effective parameters (see \autoref{sec:gf} and \autoref{sec:ep:comparison} for details).}
\label{fig:sketch_complex_E}
\end{figure}

Here, we demonstrate how \glspl{ep} naturally emerge in \gls{ll} Theory by bosonizing and studying a Hermitian microscopic one-dimensional lattice model of fermions with sublattice-dependent interactions.
Upon linearizing the dispersion relation of an extended critical \gls{ssh} model \cite{SSH_Su_1979,SSH_review_Heeger_1988} (see \autoref{fig:sketch_complex_E}a), bosonization leads to a novel \acrlong{sg}-type interaction term of the form $\cos\kla*{2\phi(x)}\partial_x\phi(x\pm a)$, where $a$ is the lattice spacing.
We compute the single-particle \gls{gf} at finite temperature by treating this term in Matsubara \gls{pt} to leading order.
Note that in our construction it is necessary to keep the lattice spacing $a$ finite to obtain rigorous results.
This gives rise to an intricate residue structure, and we handle the relating technical challenges by breaking the problem down into manageable parts.

Analytic continuation to real frequencies and inference of an effective \gls{nh} Hamiltonian from the \gls{gf} then yields quasi-particle spectra that feature \glspl{ep} (see \autoref{fig:sketch_complex_E}b) in an extended parameter range.
We demonstrate these properties by numerically implementing our perturbative solution along with an analytical continuity argument exploiting the \gls{cs} of the model.
We quantitatively analyze how the four important parameters, inverse temperature $\beta$, Luttinger parameter $K$, renormalized Fermi velocity $v$ and effective coupling $g$, continuously impact the \glspl{ep}. 
To corroborate our findings, we also provide comparisons to complementary methods based on plain fermionic \gls{pt}, as well as a fully numerical approach via \gls{csba}.

\paragraph*{Outline.}
In \autoref{sec:methods}, we introduce and bosonize our model.
The single-particle \gls{gf} is perturbatively computed in \autoref{sec:gf}.
The \gls{nh} topology of the corresponding effective Hamiltonian is subsequently discussed in \autoref{sec:ep}, where we also provide quantitative comparisons to results from fermionic \gls{pt} and microscopic numerics via \gls{csba}.
In \autoref{sec:conclusion}, we conclude our results and provide an outlook.

\section{Model and Low-Energy Theory}\label{sec:methods}
In \autoref{sec:fermionic_model}, we introduce the microscopic lattice model which we subsequently bosonize in \autoref{sec:boson}.
\subsection{Fermionic lattice model}\label{sec:fermionic_model}
We consider an extended \gls{ssh} model given by \cite{EP_DynamicInduced_Quenched_Lehmann_2021}
\begin{align}
H_0 &= -\frac{1}{2} \sum_{j=0}^{N-1} \klb*{J \textbf{c}_j^\dagger \sigma_x \textbf{c}_j - \textbf{c}_j^\dagger \kla*{t_x\sigma_x - \i t_y \sigma_y} \textbf{c}_{j+1}} + \hc,
\label{eqn:free_model}
\end{align}
where $N$ is the number of sites and $J,t_x,t_y$ are real-valued hopping parameters, and we focus on the gapless regime $J=t_x$.
The two-component annihilation operators $\textbf{c}_j = (c_{j,\A}, c_{j,\B})^\transpose$ act on site $j$, and the indices $\A,\B$ refer to the respective sublattice.
Note that $\textbf{c}_{N} \equiv \textbf{c}_0$, i.e. we assume periodic boundary conditions.
We introduce Fourier modes via $\textbf{c}_j = \frac{1}{\sqrt{N}} \sum_k \powe{\i kja} \textbf{c}_k$, where $a$ is the lattice spacing and the sum runs over momenta $k=\frac{2\pi}{Na} n_k$ with $n_k=-\frac{N}{2},\dots,\frac{N}{2}-1$ for even $N$.
Linearizing the dispersion around the Fermi energy yields
\begin{align*}
H_0 &= \sum_k \textbf{c}_k^\dagger \vf k \sigma_y \textbf{c}_k,
\end{align*}
where $\vf=t_ya$ is the Fermi velocity.
To model interactions between the particles, we include a sublattice-dependent density-density term given by \cite{EP_DynamicInduced_Quenched_Lehmann_2021,NHT_SymmetryProtectedExceptRings_Chiral_Yoshida_2019,NH_SkinEffect_QuasiParticles_NNint_Micallo_2023}
\begin{align}
H_\tint &= \sum_{\bra*{j,j'}} \sum_{s\in \klc*{\A,\B}} U_s \varnormord{n_{j,s}} \varnormord{n_{j',s}}.
\label{eqn:hint}
\end{align}
Here, $n_{j,s} = c_{j,s}^\dagger c_{j,s}$ is the density operator, $\bra*{j,j'}$ denotes nearest neighbors and $\varnormord{A} = A - \bra*{A}_0$.
We assume half-filling, thus $\bra*{n_{j}}_0 = 1$ due to translational invariance (one particle per unit cell).
The free Hamiltonian also admits sublattice symmetry, $\klc*{H_0, \sigma_z}=0$, and therefore $\bra*{n_{j,s}}_0 = \frac{1}{2}$ as in \refcite{EP_DynamicInduced_Quenched_Lehmann_2021}.

Note that the interaction strengths $U_\A$ and $U_\B$ on the sublattices are allowed to be different, which is a key requirement for \glspl{ep}, as it introduces sublattice staggering in the scattering rates.
Furthermore, one requirement for \glspl{ep} is a non-trivial matrix structure of the self-energy $\Sigma$, i.e. $\klb*{\Sigma, H_0}\neq 0$ \cite{EP_DynamicInduced_Quenched_Lehmann_2021,EP_Physics_Heiss_2012,NH_SkinEffect_QuasiParticles_NNint_Micallo_2023}. 
For $U_\A\neq U_\B$ this should be satisfied, because the interaction term facilitates a $\sigma_z$-contribution, while $H_0$ corresponds to $\sigma_y$.

\autoref{fig:sketch_complex_E} illustrates the dispersion relation with and without interactions.
In concrete examples, such as the above, we always use the parameter values $a=J=t_x=1$ and $t_y=\frac{1}{2}$.

\subsection{Bosonization}\label{sec:boson}
Before applying bosonization, we first diagonalize $H_0$ by introducing left- and right-movers $c_\ell = \sum_{s={\A,\B}} U_{\ell,s} c_s$ with $U=\frac{1}{\sqrt{2}} \begin{psmallmatrix} 1 & -\i \\ -\i & 1 \end{psmallmatrix}$, which gives \footnote{This is essentially the kinetic part of the Tomonaga-Luttinger model \cite{TLL_Mattis_1965,LL_TLL_review_Schonhammer_2003}.} (we identify $\R=+1$ and $\L=-1$, as well as $\A=+1$ and $\B=-1$)
\begin{align*}
H_0 &= \sum_{\ell=\R,\L} \sum_k \ell \vf k \varnormord{c_{k,\ell}^\dagger c_{k,\ell}}.
\end{align*}
With $n_s = \frac{1}{2} \sum_\ell \kla[\big]{n_\ell +\i\ell s c_\ell^\dagger c_{-\ell}}$ and $U_\pm = \frac{1}{2} \kla*{U_\A \pm U_\B}$ the interaction can be split into
\begin{align*}
H_+ &= \frac{U_+}{2} \sum_{\bra*{j,j'}} \sum_{\ell,\ell'} \bsplit{\kla[\Big]{&\varnormord{n_{j,\ell}} \varnormord{n_{j',\ell'}} \\
&-\ell\ell' \varnormord{c_{j,\ell}^\dagger c_{j,-\ell}} \varnormord{c_{j',\ell'}^\dagger c_{j',-\ell'}}},}\\
H_- &= U_- \sum_{\bra*{j,j'}} \sum_{\ell,\ell'} \i\ell \varnormord{c_{j,\ell}^\dagger c_{j,-\ell}} \varnormord{n_{j',\ell'}}.
\end{align*}
Using the replacement $c_{j,\ell} = \sqrt{a} \psi_\ell(ja)$ \cite{MajoranaZeroModes_PeriodicallyGated_Malard_2015}, we write $H_\pm = \frac{U_\pm a}{2} \sum_{\ell,\ell'} \intx{ }{ }{\mathcal{H}_{\pm,\ell,\ell'}}{x}$ in the spirit of a continuum limit.
The bosonization identity reads
\begin{align}
\psi_\ell(x) &= \frac{\eta_\ell}{\sqrt{L}} \powe{\i\ell \frac{2\pi}{L} N_\ell x} \normord{\powe{-\i\varphi_\ell(x)}}, \label{eqn:boson_identity}
\end{align}
where $\normord{(\dots)}$ denotes normal-ordering with respect to the bosonic operators $b_{q,\ell} = \frac{-\i}{\sqrt{n_q}} \sum_k c_{k-q,\ell}^\dagger c_{k,\ell}$ \cite[\ssec 10.A]{Bosonization_DelftSchoeller_1998}.
The bosonic fields are given by $\varphi_\ell=\varphi_\ell\eminus + \varphi_\ell\eplus$ with
\begin{align*}
\varphi_\ell\eminus(x) &= -\sum_{q>0} \frac{1}{\sqrt{n_q}} \powe{-\alpha q/2} \powe{\i q\ell x} b_{q,\ell}
\end{align*}
and $\varphi_\ell\eplus = \kla*{\varphi_\ell\eminus}^\dagger$.
Ideally, the \acrshort{uv} regulator $\alpha$ should be taken to zero, but one may argue that it can also mimic a finite bandwidth \cite[\ssec 5]{Bosonization_DelftSchoeller_1998}.
However, this leads to a subtle issue with fundamental properties of the \glspl{gf} that we discuss in \autoref{sec:gf}.

For $\ell=\L$ the dispersion is decreasing -- thus we have to count the momenta in the opposite way -- which amounts to the replacement $x \rightarrow -x$ \cite[\ssec 10.A]{Bosonization_DelftSchoeller_1998}.
It is more convenient to adjust the relations impacted by undoing this change via $\varphi_\L(-x)\rightarrow \varphi_\L(x)$, which leads to additional factors of $\ell$ (e.g. the exponent of $\powe{\i q\ell x}$).

In the thermodynamic limit, the Klein factors $\eta_\ell$ become \thermit{} \cite[\spage 25]{Bosonization_LL_LectureNotes_Schulz_1998} and we may neglect the additional phase factor in \autoref{eqn:boson_identity}.
The density operator and standard commutation relations are then given by
\begin{align*}
\rho_\ell(x) = \lim_{\epsilon\rightarrow 0} \varnormord{\psi_\ell(x)^\dagger \psi_\ell(x+\epsilon)} =  \frac{-\ell}{2\pi} \partial_x\varphi_\ell(x),\\
\klb*{\varphi_\ell\eplus(x), \varphi_{\ell'}\eminus(x')} = \delta_{\ell,\ell'} \log\klb*{\frac{2\pi}{L} \kla*{\alpha + \i\ell (x-x')}}.
\end{align*}
Our model takes the form $H=H_0+H_2+H_4$, where (see \appref{sec:app:boson} for details)
\begin{align*}
H_0 &= \frac{v}{2\pi} \intx{ }{ }{\normord{\klb*{K \kla*{\partial_x\theta}^2 + \frac{1}{K} \kla*{\partial_x\phi}^2}}}{x}
\end{align*}
corresponds to an interacting \gls{ll} with Luttinger parameter $K=\klb[\big]{1 + \frac{2U_+a}{\pi\vf} \frac{\alpha^2}{a^2+\alpha^2}}^{-1/2}$, renormalized Fermi velocity $v=\frac{\vf}{K}$ and the fields $\varphi_\ell \equiv \ell \phi - \theta$.
The remaining perturbations read
\begin{align*}
H_2 &= \frac{U_-a}{4\pi^2\alpha} \i\eta_\L \eta_\R \sum_\ell \sum_{s_a=\pm 1} \intx{ }{ }{\powe{2\i\ell \phi(x)} \partial_x\phi(x+s_aa)}{x},\\
H_4 &= \frac{U_+a}{4\pi^2\alpha^2} \sum_\ell \intx{ }{ }{\powe{4\i\ell \phi(x)}}{x},
\end{align*}
and we refer to $H$ as the \gls{ell}.
Note that $H$ itself is still \thermit{} and that it is not straightforward to take either limit $a,\alpha\rightarrow 0$, as this would give a trivial $K=1$ and eliminate $H_2$ due to its integrand becoming a total derivative.

\section{Perturbative Approach}\label{sec:gf}
In this section, we are interested in the single-particle Matsubara \acrlong{gf} that in fermionic and bosonic language is given by (imaginary time $\tau = \i t$)
\begin{align*}
G^\M_{\ell,\ell'}(x,\tau) &= -\bra*{\mathcal{T}_\tau \psi_\ell(x,\tau) \psi_{\ell'}^\dagger (0,0)} \\
&= -\frac{\eta_\ell \eta_{\ell'}}{2\pi\alpha}\bra*{\mathcal{T}_\tau \powe{-\i\varphi_\ell(x,\tau)} \powe{\i\varphi_{\ell'}(0,0)}}.
\end{align*}
Due to the complex (non-bilinear) nature of the model, an exact evaluation of the expectation value is infeasible, and we instead resort to a perturbative approach.
We start by computing the free \gls{gf} in \autoref{sec:order0}, before discussing the first order corrections in \autoref{sec:order1}.

\subsection{Zeroth order}\label{sec:order0}
We intentionally write $H$ in a form that is not normal-ordered, because one can evaluate the expectation value of an arbitrary product of vertex operators,
\begin{align*}
\bsplit{\bra[\Big]{\prod_j &\powe{\i \klb*{A_j \phi(z_j) + B_j\theta(z_j)}}}_0 = \\
&\prod_{j<k} \prod_\ell \powe{-\frac{1}{4} \klb*{KA_jA_k + \frac{1}{K} B_jB_k + \ell \kla*{A_jB_k + A_kB_j}} \Phi(z_{jk,\ell})}.}
\end{align*}
Here, we introduce $z=(z_{+}, z_{-})$ with the dimensionless variables $z_{\pm}=\frac{\pi}{\beta v} \kla*{v\tau \pm \i x}$ for each space-time point $(x_j,\tau_j)$.
Furthermore, we write $z_{jk} = z_j-z_k$ and assume the neutrality conditions $\sum_jA_j = 0$ and $\sum_j B_j=0$; otherwise the right hand side is zero \cites[\ssec 3]{Gogolin_1999}[\sapp C]{Giamarchi_2004}.
The function $\Phi(z) = \bra{\varphi_\R(x,\tau)\varphi_\R(0,0) - \varphi_\R(0,0)^2}_0$ is given by (see \appref{sec:app:special_func})
\begin{align*}
\log\klb*{\frac{\Gamma\kla[\big]{1 + \frac{w + z_+}{\pi}} \Gamma\kla[\big]{1 + \frac{w - z_+}{\pi}}}{\Gamma\kla[\big]{1 + \frac{w}{\pi}}^2 \kla*{1 + \frac{z_+}{w}}}} \overset{w\ll |z_+|}{\simeq} \log\klb*{w \csc(z_+)},
\end{align*}
where $w=\frac{\pi\alpha}{\beta v}$.
Interestingly, although the condition for the asymptotic equality is not always satisfied, the approximation of $\Phi(z)$ appears to be the physically correct result.
To see why, consider the single-particle Matsubara \gls{gf} for $H_0$,
\begin{align*}
G_{0,\ell,\ell'}^\M(z) &= -\bra*{\mathcal{T}_\tau \psi_\ell(z) \psi_{\ell'}^\dagger(0)}_0 \\
&= -\frac{\delta_{\ell,\ell'}}{2\pi\alpha} \prod_{s=\pm 1} \powe{-\frac{1}{4} \klb*{-K - \frac{1}{K} + 2\ell s} \Phi(z_s)}\\
&\asymeq{\alpha\rightarrow 0} -\frac{\delta_{\ell,\ell'} w^{2M}}{2\beta v} \csc\kla*{z_{-\ell}}^{M+1} \csc\kla*{z_{\ell}}^M, \numberthis\label{eqn:gf_order0_matsubara_real_space}
\end{align*}
where $M=\frac{1}{4} \klb*{K + \frac{1}{K} - 2}\ge 0$.
This result is consistent with the generic anti-periodicity of fermionic Matsubara \glspl{gf}, but only within the asymptotic limit $\alpha\rightarrow 0$, i.e. for the approximate form of $\Phi(z)$.
Nevertheless, this matches the result stated in literature, e.g. \cites[\ssec IV.C]{Bosonization_FiniteT_Bowen_2001}[\ssec III.A]{LL_Electron_Dephasing_Gornyi_2005}{LL_Electron_Lifetime_LeHur_2006}[\ssec 6.8.4]{Fradkin_2013}.

The \gls{ft} of $G_0$ (see \appref{sec:app:detailed_ell_gf} for details) leads to
\begin{align*}
G_{0,\ell,\ell'}^\M(k,\i\omega_n^\F) &= -\delta_{\ell,\ell'} \frac{\beta w^{2M}}{2\pi^2} J_{M,1}\kla*{-\ell \frac{\beta vk}{\pi}, \frac{\beta\i\omega_n^\F}{\pi}},
\end{align*}
where $\omega_n^\F = \frac{(2n+1)\pi}{\beta}$ for $n\in\mathbb{Z}$ are the fermionic Matsubara frequencies and the functions $J_{b,N}$ are defined in \appref{sec:app:special_func}.
As a side note, the \gls{ft} of $G_0^\ret(k,\omega)$ to $(k,t)$ via a straightforward application of the residue theorem yields the lifetime $\tau = \frac{\beta}{2\pi M}$ in accordance with \cite{LL_Electron_Lifetime_LeHur_2006}.

\subsection{Leading corrections}\label{sec:order1}
For now, we ignore $H_4$ and focus on the perturbation $\sact_\tint = \intx{0}{\beta}{H_2}{\tau}$.
Then the first order correction is
\begin{align*}
&G^\M_{1,\ell,\ell'}(z) = +\bra*{\mathcal{T}_\tau \psi_\ell(z) \psi_{\ell'}^\dagger(0) \sact_\tint }_0\\
&\quad = \sum_{s,s_a=\pm 1} \frac{\i sv g}{2\pi\alpha^2} \int \bsplit{\bra[\Big]{&\mathcal{T}_\tau \eta_\ell \powe{-\i\varphi_\ell(z)} \eta_{\ell'} \powe{\i\varphi_{\ell'}(0)} \eta_{-s} \eta_{s}\\
&\times \powe{2\i s\phi(z')} s_a\partial_a (-\i\partial_J) \powe{\i J\phi(\tilde{z}')}}_0,}
\end{align*}
where $g=\frac{U_-a}{4\pi^2v}$, $\tilde{z}=(\tilde{z}_+,\tilde{z}_-)$ corresponds to $z$ but with shifted spatial coordinate $\tilde{x}=x+s_aa$, and $J\rightarrow 0$ is implied.
The neutrality violating factor drops out due to $\partial_J \powe{AJ^2 + BJ} = B$, and the neutrality condition requires $\ell'=-\ell$ and $s=\ell$.
The time-ordering splits the integration over $\tau'$, but one can show that it is possible to recombine the two parts by carefully treating sign factors.
Thus we find
\begin{align*}
\int &\sum_{s,s_a=\pm 1} \frac{-\i\ell\pi (1-Ks)}{4\beta v} \frac{\pi v g\delta_{\ell,-\ell'} w^{2M+2K_-}}{2(\beta v)^2} \\
&\times \csc(z_+)^{M-K_-} \csc(z_-)^{M-K_-} \csc(z_{-\ell}')^{K_-} \csc(z_{\ell}')^{K_+} \\
&\times \csc(z_{-\ell}-z_{-\ell}')^{K_+} \csc(z_{\ell}-z_{\ell}')^{K_-} \\
&\times \klb*{\cot\kla*{z_{\ell s}-z_{\ell s}' - \i \ell ss_a \tilde{a}} + \cot\kla*{z_{-\ell s}' - \i \ell ss_a\tilde{a}}}
\end{align*}
with $\tilde{a} = \frac{\pi a}{\beta v}$ and $K_\pm = \frac{K\pm 1}{2}$.
The \gls{ft} can be performed via the \gls{ct} and leads to $G_1^\M(k,\i\omega_n^\F) = \beta g w^{2M+K-1}\sigma_y \tilde{G}_1^\M$, where
\begin{align*}
\tilde{G}_1^\M &= \int_\mathbb{R} \sum_{s,s'=\pm 1} \sum_{m\in\mathbb{Z}} \frac{(1+Ks)}{8\pi^4} J_{M-K_-,0}\kla*{k',2\i m} \\
&\qquad \times J_{K_-,1} \kla*{s' \frac{\beta v}{\pi}k-s' k',\frac{\beta\i\omega_n^\F}{\pi} - 2\i m} \\
&\qquad \times S_{K_-,1,s}\kla*{s' k' - s' \frac{\beta v}{\pi} k,\frac{\beta\i\omega_n^\F}{\pi} - 2\i m, \tilde{a}}\,\d k'
\end{align*}
with $S_{b,1,s}$ defined in \appref{sec:app:special_func}.
Performing the analytic continuation $\i\omega_n^\F\rightarrow \omega + \i\eta$ is now straightforward and yields the retarded \gls{gf} to first order in $H_2$.
Note that we identify the ratio
\begin{align}
\PTR(k,\omega) &= \abs*{g\frac{G_1(k,\omega)}{G_0(k,\omega)}} \label{eqn:ell:gf:PTR_def}
\end{align}
as the ``small quantity'' of the perturbation series, and $R\approx 1$ coincides with several breakdowns of the theory, e.g. positive imaginary energies and unphysical divergences.

We can justify disregarding the perturbation $H_4$ for the calculation of the \gls{gf} via a \gls{rg} argument.
For this purpose, we use a standard perturbative momentum-shell \gls{rg} \cite{Bosonization_Shankar_1994,AltlandSimons,RG_NonPerturbative_Introduction_Delamotte_2012,Fradkin_2013}.
Consistent with power counting, the tree-level flow equations are
\begin{align*}
\frac{\d g_4}{\d l} &= \kla*{2 - 4K} g_4 \quad\text{and}\quad \frac{\d g_2}{\d l} = \kla*{1 - K} g_2,
\end{align*}
where at \gls{rg}-time $l=0$ we have $g_4 = \frac{U_+a}{4\pi^2v}$ and $g_2 = \frac{U_-a}{4\pi^2v}$.
Thus, $H_4$ is indeed irrelevant for $K>\frac{1}{2}$ (which is typically satisfied, see \autoref{sec:ep:comparison}).
In \appref{sec:app:rg}, we also derive the one-loop corrections for an arbitrary cut-off function.

\section{Exceptional Points}\label{sec:ep}
Using the single-particle \gls{gf}, the effective Hamiltonian can be defined as $H_\eff^\ret(k,\omega) = (\omega+\i\eta) - \kla*{G^\ret(k,\omega)}^{-1}$, or, in terms of the self-energy, as $H_\eff^\ret(k,\omega) = H_0(k) + \Sigma^\ret(k,\omega)$ \cite{NHT_FiniteLifetime_QuasiParticles_Kozii_2017,NH_Heff_SelfEnergy_1d_Hubbard_Rausch_2021,GF_SelfEnergyFunctionals_Eder_2021}.
In \autoref{sec:ep:symmetry_heff}, we discuss our results by analysing the quasi-particle spectrum of $H_\eff$ with a focus on the occurrence and stability of \glspl{ep}.
We discuss the quantitative dependence of the characteristic parameters of the \glspl{ep} in \autoref{sec:ep:eps_par_dependence} and compare to \gls{csba} as well as fermionic plain \gls{pt} in \autoref{sec:ep:comparison}.

\subsection{Symmetries of the effective Hamiltonian}\label{sec:ep:symmetry_heff}
In the following, quantities are always assumed to be in the $\A\B$-basis, i.e. indices correspond to a sublattice.
Whenever an object is instead in the basis of left- and right-movers, we denote it with a subscript.

Our result for the single-particle \gls{gf} takes the form
\begin{align*}
G^\ret_{\R\L}(k,\omega) &= \begin{pmatrix}
G_0(k,\omega) & -\i gG_1(k,\omega) \\ \i gG_1(k,\omega) & G_0(-k,\omega)
\end{pmatrix} + \mathcal{O}(g^2),
\end{align*}
where $G_0\equiv G_{0,\R,\R}^\ret$ and $G_1 \equiv \frac{\i}{g} G_{1,\R,\L}^\ret \equiv \beta w^{2M+K-1} \tilde{G}_1^\M$.
Writing $G_\pm(k) = \frac{1}{2} \klb*{G_0(k) \pm G_0(-k)}$, we have $G^\ret_{\R\L} = G_+ \sigma_0 + \textbf{g}_{\R\L}\cdot \bm{\sigma}$.
Here, $\bm{\sigma}$ is the vector of Pauli matrices, $\textbf{g}_{\R\L} = (0, gG_1, G_-)^\transpose$, and by construction $G_+$ ($G_-$) is symmetric (anti-symmetric) in $k$.

Our model admits chiral symmetry $\chs$ with the transformation matrix $U_S=\sigma_z$ in the $\A\B$-basis.
The important consequence of this symmetry is the relation \cite{EP_DynamicInduced_Quenched_Lehmann_2021} (see \appref{sec:app:cs} for a derivation)
\begin{align*}
\sigma_z H_\eff^\dagger(k,\omega) \sigma_z &= -H_\eff(k,-\omega)
\end{align*}
for the effective Hamiltonian defined via the single-particle \gls{gf} $\sgf^\ret(k,\omega) = \kla*{\omega + \i \eta - H_\eff(k,\omega)}^{-1}$.
Due to the $2\times 2$-matrix structure we can write $H_\eff = d_0 \sigma_0 + \textbf{d}\cdot\bm{\sigma}$ (arguments $k$ and $\omega$ are omitted), where $d_0\in\mathbb{C}$ and $\textbf{d}=\textbf{d}_\r + \i \textbf{d}_\i$ with $\textbf{d}_\r, \textbf{d}_\i \in\mathbb{R}^3$.
The conditions for an \gls{ep} then become \cite{ExceptTop_RevModPhys.93.015005}
\begin{align}
\textbf{d}_\r^2 &= \textbf{d}_\i^2 \qquad\text{and}\qquad \textbf{d}_\r\cdot \textbf{d}_\i = 0, \label{eqn:dr2_di2_and_drdi_condition}
\end{align}
and in one dimension stable \glspl{ep} generically require an additional symmetry to reduce the codimension of these equations:
In our case, the chiral symmetry trivializes $\textbf{d}_\r\cdot\textbf{d}_\i = 0$ at the Fermi energy $\omega=0$ \cite{NHT_SymmetryProtectedNodalPhases_Budich_2019}.

Let us now explicitly compute the coefficients and demonstrate the occurrence of \gls{ep}.
With $U$ from \autoref{sec:boson},
\begin{align*}
G^\ret &= U G^\ret_{\R\L} U^\dagger = G_+\sigma_0 + \textbf{g} \cdot \bm{\sigma}
\end{align*}
where $\textbf{g}=(0, -G_-, gG_1)^\transpose$.
Omitting the frequency argument, the functions $d_0$ and $\textbf{d}$ are then given by \footnote{Note that we have the general result $\kla*{g_0 \sigma_0 + \textbf{g}\cdot \bm{\sigma}}^{-1} = \frac{1}{g_0^2 - \textbf{g}^2} \kla*{g_0 \sigma_0 - \textbf{g}\cdot\bm{\sigma}}$ for non-singular matrices. }
\begin{align*}
d_0(k) &= \frac{\omega - G_+(k)}{G_0(k) G_0(-k) - g^2G_1(k)^2}, \\
\textbf{d}(k) &= -\frac{\textbf{g}(k)}{G_0(k) G_0(-k) - g^2G_1(k)^2}.
\end{align*}
Since $G_1(k,\omega)=G_1(-k,\omega)$, this means that $d_0$ and $d_z$ are symmetric in $k$, while $d_y$ is anti-symmetric.
Due to $G_0(k,0)=-G_0(-k,0)^\ast$, we have $G_-(k,0) = \Re G_0(k,0)$.
Similarly $J_{b,0}(k,\i\omega)\in\mathbb{R}$, $J_{b,1}(k,\i\omega)^\ast = -J_{b,1}(-k,\i\omega)$ and $S_{b,1,s}(k,\i\omega,a)^\ast = S_{b,1,s}(-k,\i\omega,a)$ lead to $G_1(k,0)^\ast = -G_1(k,0)$, and thus the remaining condition for an \gls{ep} simplifies to $\abs*{G_-(k_\text{EP},0)} = \abs*{g G_1(k_\text{EP},0)}$.
By construction $G_-(0,\omega)=0$, so $\abs*{G_-(k,0)}$ must be increasing on some interval $k\in (0,k_1)$.
Fixing such a $k$, we then have $\abs*{G_-(k,0)}>\abs*{g G_1(k,0)}$ for sufficiently small $g$, and consequently an \gls{ep} must exist by virtue of the intermediate value theorem (note that $G_1(0,0)\neq 0$ because the imaginary part of the $k'$-integrand is symmetric).
 
\subsection{EPs and parameter dependence}\label{sec:ep:eps_par_dependence}
The complicated structure of the formal solution makes quantitative comparison to fermionic \gls{pt} and \gls{csba} difficult.
We therefore proceed to numerically explicate our formal solution for the \gls{gf}.

\begin{figure}[htp]
\centering
\includegraphics{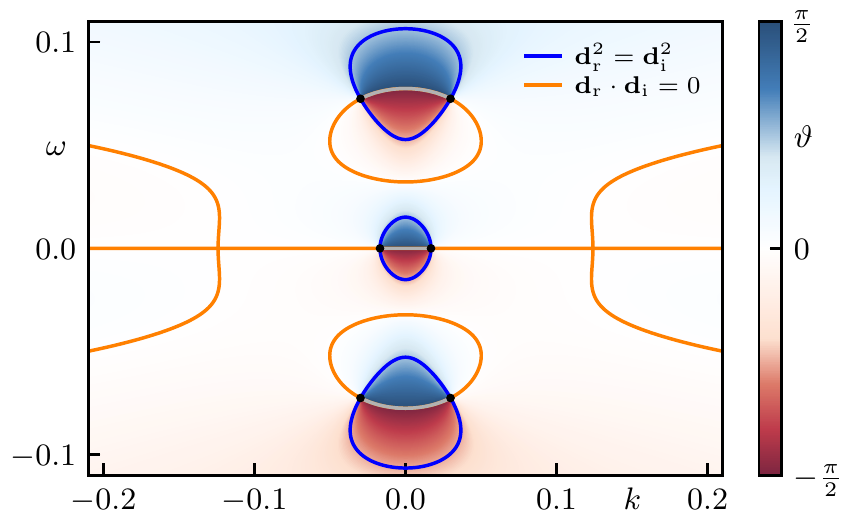}
\caption{Visualization of the contour lines for \autoref{eqn:dr2_di2_and_drdi_condition} as well as the phase of the complex energy gap $\vartheta$ in frequency space.
\glspl{ep} correspond to intersections of the contours and are marked with black dots connected by imaginary Fermi arcs \cite[\ssec II.B]{ExceptTop_RevModPhys.93.015005}.
The parameters are $\beta=20$, $K=0.4$, $v=0.8$, $g=0.1$ and $\alpha=2$ ($\PTR_\text{max} \approx 0.3$).}
\label{fig:ep_contour_dual_example}
\end{figure}
The energies are $E_\pm = d_0 \pm \sqrt{\textbf{d}_\r^2 - \textbf{d}_\i^2 + 2\i\textbf{d}_\r \cdot \textbf{d}_\i}$ and $\vartheta = \arg\kla*{E_+-E_-}$ is the phase of the energy gap; \autoref{fig:ep_contour_dual_example} illustrates our \gls{ell} result for a specific choice of parameters.
Note that qualitatively this is remarkably similar to the brute force numerical results in \refcite{EP_DynamicInduced_Quenched_Lehmann_2021}: Our result shares all the symmetries and features, such as the three arcs protruding from the horizontal line, and the additional pair of \glspl{ep} at finite $\omega$.
The main visual difference is that, in the \gls{csba} result, the three lobes of $\textbf{d}_\r^2 = \textbf{d}_\i^2$ are connected into one contour.
However, this is merely a quantitative problem, and one should keep in mind that the temperatures are very different.

\begin{figure}[!ht]
\centering
\includegraphics{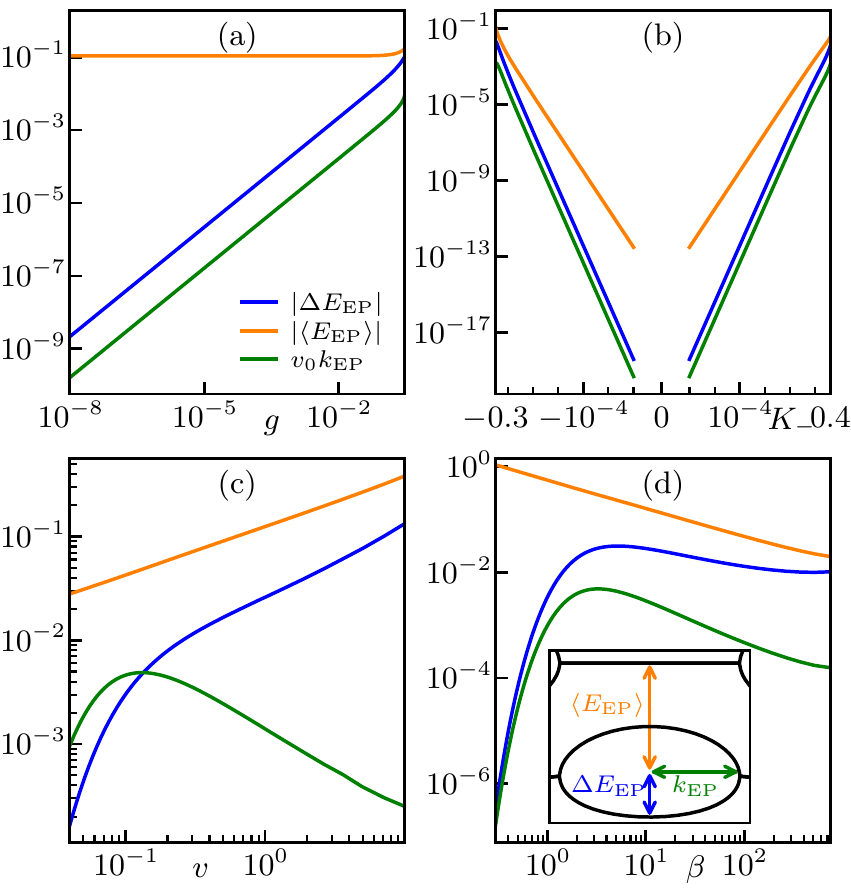}
\caption{Plot of the size and offset of the \glspl{ep}, as well as the location of the \gls{ep} (scaled by $v_0=0.1$ to give an energy) for $\beta=20$, $K=0.4$, $v=0.8$ and $g=0.1$ ($\alpha=2$).
The inset in (d) visualizes the meaning of the three variables.
In panel (a), $g$ is varied, and as anticipated from a first order calculation the size of the gap scales linear with the coupling (and so does $k_\text{EP}$).
The deviation from this behavior toward larger couplings corresponds to the breakdown of the first order approximation and coincides with $\PTR\approx 1$.
In panel (b), $K=1+2K_-$ is varied, and the $x$-axis is symmetrically logarithmic around $K=1$ (hence the visual gap around $K_-=0$).
The advantage of this visualization is that the symmetry $K\leftrightarrow \frac{1}{K}$ becomes apparent.
As in (a), the offset is primarily dictated by the free \gls{gf} while the size and location of the \glspl{ep} are determined by the perturbation.
In panel (c), $v$ is varied, and here the impact on size and location is asymmetric, thus the velocity changes the shape of the ``bubble''.
Finally, in (d) $\beta$ is varied, and as expected the \glspl{ep} vanish for sufficiently large or small temperatures.}
\label{fig:ep_size_quad_example}
\end{figure}

\autoref{fig:ep_size_quad_example} illustrates the dependence of the \gls{ep} on the four parameters $g$, $K$, $v$ and $\beta$.
Fermi liquid theory breaks down in one dimension, because the lifetime asymptotically scales as $\tau \sim \beta$ \cite{LL_Electron_Lifetime_LeHur_2006}, and thus does not diverge fast enough for well-defined quasi-particles.
Yet, it is still natural to expect any imaginary parts at $\omega = 0$ to vanish in the zero temperature limit.
Panel (d) demonstrates this, as the \glspl{ep} shrink as $\beta\rightarrow \infty$, albeit slowly ($\bra*{E_\text{EP}} \sim \beta^{2M-1}$ for $g=0$, slightly modified for finite $g$).

\subsection{Comparison to other methods}\label{sec:ep:comparison}
In this section, we compare our first order \gls{ell} \gls{pt} to fermionic second order plain \gls{pt} (see \appref{sec:app:fermionic_pt}), as well as to the numerics on the full lattice model within \gls{csba} along the lines of \refcite{EP_DynamicInduced_Quenched_Lehmann_2021}.
This comparison is challenging, because the regions in which the three methods are valid are almost disjoint: The \gls{ell} \gls{pt} is limited to comparatively low temperatures $\beta \gtrsim 1$, while the range of validity is $\beta \lesssim 1$ for the fermionic \gls{pt}.
\gls{csba} requires interactions strengths $|U_\A| \gtrsim 0.1$ to make interaction effects visible at the feasible system sizes, which is roughly the upper bound for the fermionic \gls{pt}.

\begin{figure}[htp]
\centering
\includegraphics{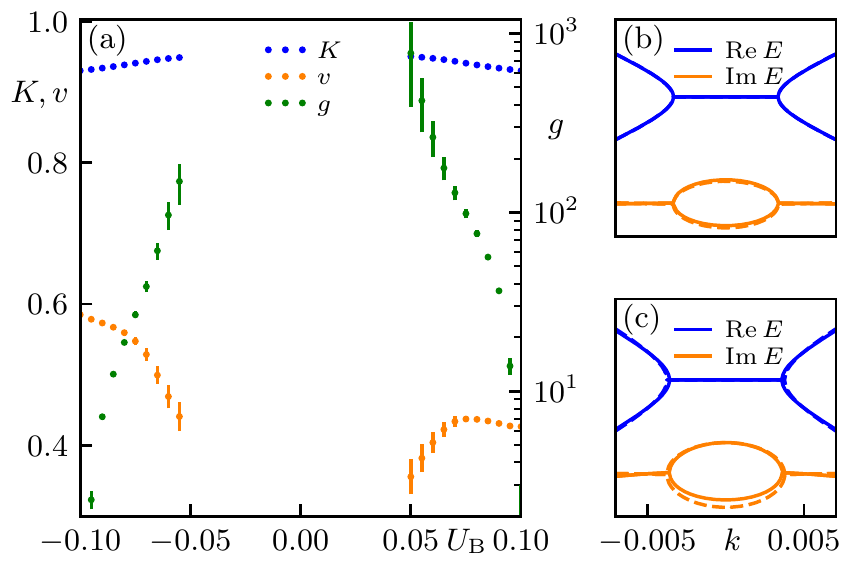}
\caption{Panel (a) illustrates optimal parameters $K$, $v$ and $g$ for a fit of the \gls{ell} \gls{pt} to the fermionic \gls{pt} at $\beta=1$ and $U_\A=0.1$ as a function of $U_\B$. 
The coupling shows exponential dependence on $U_\B$ and is therefore measured on the logarithmic axis to the right of (a).
Two examples of the result of the fit are shown in panel (b) and (c) where $U_\B=0.08$ and $U_\B=-0.07$ respectively (dashed line for fermionic \gls{pt}).}
\label{fig:ll_fpt_fit}
\end{figure}

\begin{figure}[htp]
\centering
\includegraphics{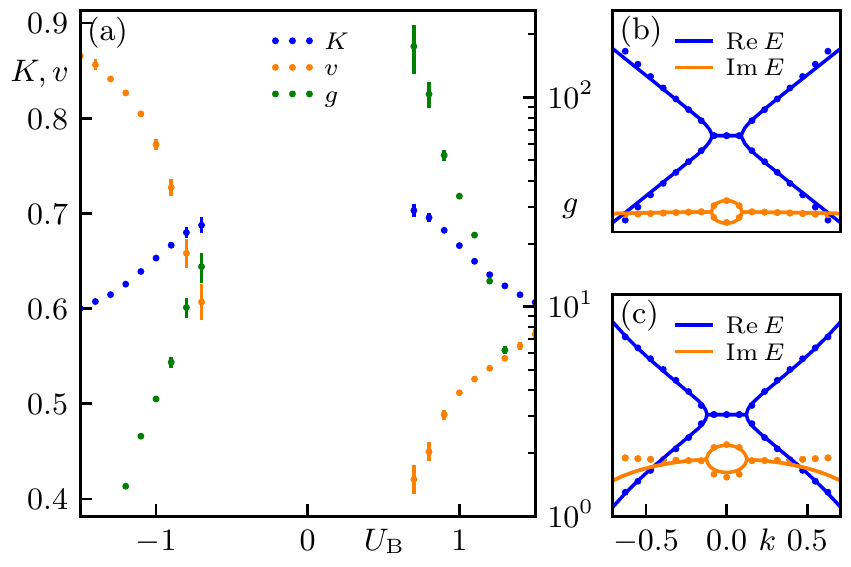}
\caption{Counterpart to \autoref{fig:ll_fpt_fit}, now comparing the \gls{ell} \gls{pt} to the \gls{csba} result at $U_\A=1.5$.
Two examples of the fit are shown in panel (b) and (c) where $U_\B=1.2$ and $U_\B=-0.9$ respectively (dots for \gls{csba}).
Note that the gap in (a) around $U_\B=0$ is not accessible using the \gls{ell} \gls{pt} and the underlying problem is exemplified in (c) by the onset of a divergent imaginary part toward larger $k$.}
\label{fig:ll_num_fit}
\end{figure}

Using the explicit formulas for $K$, $v$ and $g$ from \autoref{sec:boson} does not yield quantitative agreement, and we instead use them as fit parameters to demonstrate that the \gls{ell} \gls{pt} can in principle reproduce the results from the other methods.
The hope is that it might then be possible to predict these values via a \acrlong{rg} analysis; we briefly discuss this in \appref{sec:app:rg}.
The results of this fit to the fermionic \gls{pt} for $|U_\B|\le U_\A=0.1$ and \gls{csba} for $|U_\B|\le U_\A=1.5$ are shown in \autoref{fig:ll_fpt_fit} and \autoref{fig:ll_num_fit} respectively.
We once again fix $\alpha=2$, but smaller $\alpha$ merely lead to a constant shift of the values for $g$ (since $K<1$, smaller $\alpha$ give smaller $g$).
This is because the \glspl{ep} only depend on the ratio between $G_0$ and $G_1$, and a change in $\alpha$ can thus always be compensated by a change in $g$.

Note that qualitatively, the figures are almost identical.
Let us start the discussion starting from the right of panel (a), i.e. $U_\B=U_\A$.
Here, both the fermionic \gls{pt} and \gls{csba} predicts a small real-valued gap, which is not contained in the \gls{ell} result (this is a consequence of the linearization eliminating the $\sigma_x$-part).
Since there are no \glspl{ep}, $g=0$ and $K$ is determined by the imaginary part of the energy, while $v$ roughly corresponds to the slope of the linear section (this is also the case for $U_\B=-U_\A$).

As $U_\B$ is decreased, the coupling increases -- but contrary to the microscopic prediction, the increase is exponential rather than linear.
Furthermore, while at least qualitatively a decrease in $K$ is to be expected, the value of $v$ is much further from the simple $v=\frac{\vf}{K}$.
Unsurprisingly, the exponential growth of $g$ rapidly increases $\PTR$, and at around $\frac{U_\B}{U_\A} \approx \frac{1}{2}$ ($U_\B=0.05$ or $U_\B=0.7$) we can no longer obtain a reasonable fit using our first order \gls{ell} \gls{pt}.
Intuitively, approaching $U_\B=0$ is infeasible even if truncating the perturbation series at a higher order.
This is because the ``bubble'' constituting the imaginary energy gap has to touch the origin at $k=0$.
But, in the language of the \gls{ell} this scenario is highly unstable, as an arbitrarily small increase in $g$ would then immediately lead to a positive imaginary energy.
Another hint that this barrier may be insurmountable is the qualitative change in the behavior for $U_\B<0$:
Evidently, the dependence of $g$ is once again exponential, but despite the increase in $U_-$ the coupling actually goes down.
Similarly, the decrease in $U_+$ would naively lead to an increase of $K$ towards unity, while we observe the opposite.
Finally, $v$ is much larger than for $U_\B>0$ (which also compensates for the significantly smaller coupling, see \autoref{fig:ep_size_quad_example}c).

At least partially, the quantitative problems are presumably due to the comparatively high temperature of $\beta=1$.
Using the linearized instead of the exact dispersion in the fermionic \gls{pt} also leads to strong deviations, which suggests the excitation of degrees of freedom beyond the momenta where the linearization provides a good approximation.
Consequently, there is every reason to assume that the \gls{ll} approach suffers from this as well, and that the accuracy significantly improves for larger $\beta$ (e.g. as chosen for \autoref{fig:ep_contour_dual_example}).

\section{Concluding discussion}\label{sec:conclusion}

By bosonizing and perturbatively solving a microscopic lattice model with sublattice dependent interaction in the low energy limit, we have demonstrated how exceptional points naturally emerge in Luttinger liquid theory.
The essential new component is the \acrlong{sg}-type perturbation $H_2$, which at finite temperature and for a non-trivial Luttinger parameter $K\neq 1$ leads to the formation of \acrlongpl{ep}.
Qualitatively, our analytic results match predictions from numerical \gls{csba} data for the microscopic model along the lines of \cite{EP_DynamicInduced_Quenched_Lehmann_2021} as well as plain fermionic \acrlong{pt}.
The structural similarity of our \autoref{fig:ep_contour_dual_example} to Fig.~2 in \cite{EP_DynamicInduced_Quenched_Lehmann_2021} is particularly striking.
Quantitative agreement in our present analysis requires fitting parameters of the effective bosonized model to the numerical data.
It might be possible to even predict these parameter values via \gls{rg}, where non-perturbative schemes might be necessary since we work quite far from the zero temperature fixed point.

While this issue is not specific to our model, we find it unsatisfactory that, for $K\neq 1$, there is a residual dependence on the \acrshort{uv} cut-off $\alpha$ in the final result.
Although a finite value of $\alpha\propto\frac{1}{\beta v}$ can be argued via \gls{rg} (or $\alpha\approx a$ for a nearest-neighbor interaction \cite[\ssec 2.2.2]{Giamarchi_2004}), the analytic inconsistency with the fundamental periodicity of Matsubara \glspl{gf} remains.
On a similar note, we cannot quite take a proper continuum limit for the bosonized Hamiltonian, because setting $a=0$ eliminates the all-important perturbation $H_2$.
Interestingly, this problem cannot easily be solved by an expansion in the lattice spacing, as intermediate results are non-perturbative in $a$ due to the residue structure in certain integrals.
Thus, removing the mild non-locality induced by $a$ while retaining the \glspl{ep} requires additional insight.

Beyond our present study showing that \acrlong{ll} theory is capable of describing \glspl{ep}, it will be interesting to see which other \gls{nh} phenomena, such as the \gls{nh} Skin Effect or the anomalous spectral sensitivity of \gls{nh} systems can be investigated using the \acrlong{ll}  framework.

\begin{acknowledgments}
We acknowledge financial support from the German Research Foundation (DFG) through the Collaborative Research Centre (SFB 1143, project ID 247310070) and the Cluster of Excellence ct.qmat (EXC 2147, project ID 390858490).
Our numerical calculations were performed on resources at the TU Dresden Center for Information Services and High Performance Computing (ZIH).
\end{acknowledgments}

\appendix

\section{\gls{ell} bosonization details}\label{sec:app:boson}
In this section, we give some additional details of the bosonization procedure for the \gls{ell} left out in \autoref{sec:boson}.

Recall that $H_\pm = \frac{U_\pm a}{2} \sum_{\ell,\ell'} \intx{ }{ }{\mathcal{H}_{\pm,\ell,\ell'}}{x}$, where
\begin{align*} 
\mathcal{H}_{+,\ell,\ell'} &= \bsplit{\sum_{x'=x\pm a} \klb[\Big]{&\rho_\ell(x) \rho_{\ell'}(x+\epsilon) \\
&-\ell\ell' \varnormord{\psi_\ell^\dagger(x) \psi_{-\ell}(x)} \varnormord{\psi_{\ell'}^\dagger(x') \psi_{-\ell'}(x')}},}\\
\mathcal{H}_{-,\ell,\ell'} &= \sum_{x'=x\pm a}  2 \i\ell \varnormord{\psi_\ell^\dagger(x) \psi_{-\ell}(x)} \rho_{\ell'}(x').
\end{align*}
Since $\bra[\big]{c_{j,\ell}^\dagger c_{j,-\ell}}_0 = 0$, we may also drop the regularization for the corresponding terms.
\autoref{eqn:boson_identity} then leads to
\begin{align*}
\psi_\ell^\dagger(x) \psi_{-\ell}(x) &= \frac{\eta_\ell^\dagger\eta_{-\ell}}{L} \powe{\i \frac{2\pi}{L} \klb*{N_\ell - N_{-\ell} - 1}x} \normord{\powe{\i\varphi_\ell(x) - \i\varphi_{-\ell}(x)}} \\
&\overset{L\rightarrow \infty}{\sim} \frac{\eta_\ell \eta_{-\ell}}{L} \normord{\powe{\i\varphi_\ell(x) - \i\varphi_{-\ell}(x)}}.
\end{align*}
Normal-ordering the bilinear $\rho_\ell \rho_{\ell'}$ only amounts to an irrelevant additive constant, and introducing new fields via $\varphi_\ell = \ell\phi - \theta$ yields
\begin{align*}
\mathcal{H}_{+,\ell,\ell'} &= \bsplit{\sum_{x'=x\pm a} \klb[\bigg]{&\frac{\ell\ell'}{4\pi^2} \normord{\partial_x\varphi_\ell(x) \partial_x \varphi_{\ell'}(x')} \\
&+ \frac{1}{L^2} \normord{\powe{2\i\ell \phi(x)}}  \normord{\powe{2\i\ell'\phi(x')}}},}\\
\mathcal{H}_{-,\ell,\ell'} &= \sum_{x'=x\pm a} 2 \i\ell \frac{\eta_\ell \eta_{-\ell}}{L} \normord{\powe{2\i\ell\phi(x)}} \frac{-\ell'}{2\pi}\partial_x\varphi_{\ell'}(x').
\end{align*}
Here, as well as in the following, the asymptotic limit is always implied.
By using the commutation relation
\begin{align*}
\klb*{\phi\eplus(x+a), \phi\eminus(x)} &= \frac{1}{4} \sum_{\ell,\ell'} \ell\ell' \klb*{\varphi_\ell\eplus(x+a), \varphi_{\ell'}\eminus(x)} \\
&= \frac{1}{2} \log\klb*{\frac{2\pi}{L} \sqrt{\alpha^2+a^2}}
\end{align*}
we can normal-order the product of vertex operators, $ \normord{\powe{\i m\phi(x)}} \normord{\powe{\i m'\phi(x+a)}}$, by rewriting it as
\begin{align*}
\powe{\i m\phi\eplus(x)} &\powe{\i m\phi\eminus(x)} \powe{\i m'\phi\eplus(x+a)} \powe{\i m'\phi\eminus(x+a)} \\
&= \normord{\powe{\i m\phi(x) + \i m'\phi(x+a)}} \powe{\klb*{\i m\phi\eminus(x),\, \i m'\phi\eplus(x+a)}} \\
&= \normord{\powe{\i m\phi(x) + \i m'\phi(x+a)}} \klb*{\frac{2\pi}{L} \sqrt{\alpha^2+a^2} }^{\frac{mm'}{2}}.
\end{align*}
To normal-order $H_-$, first note that $\klb*{A, \powe{B}} = C\powe{B}$ for $C=\klb*{A,B} \in\mathbb{C}$ leads to
$\klb[\big]{\phi\eplus(x+a), \powe{\i m\phi\eminus(x)}} = \frac{\i m}{2} \log\klb*{\frac{2\pi}{L}\sqrt{\alpha^2+a^2}} \powe{\i m\phi\eminus(x)}.$
This commutator can then be used to rewrite $\normord{\powe{\i m\phi(x)}} \partial_x\phi\eplus(x+a)$ as
\begin{align*}
\normord{\powe{\i m\phi(x)} \partial_x \phi\eplus(x+a)} - \frac{\i m}{2} \frac{a}{a^2+\alpha^2}  \normord{\powe{\i m\phi(x)}}.
\end{align*}
The similar expression for $\phi\eminus(x+a)$ is already normal-ordered and, since we sum on $\pm a$, this leads to
\begin{align*}
\sum_{x'=x\pm a} \normord{\powe{\i m\phi(x)}} \partial_{x'} \phi(x') &= \sum_{x'=x\pm a} \normord{\powe{\i m\phi(x)}\partial_{x'}\phi(x')}.
\end{align*}
Interestingly, the product in $H_-$ therefore already is normal-ordered and thus far the full Hamiltonian is $H=H_0 + H_1 + H_2 + H_3 + H_4$, where
\begin{align*}
H_0 &= \frac{\vf}{2\pi} \int \normord{\klb*{\kla*{\partial_x\theta}^2 + \kla*{\partial_x\phi}^2}},\\
H_1 &= \frac{U_+a}{2\pi^2} \sum_{s_a=\pm 1} \int  \normord{\partial_x\phi(x) \partial_x\phi(x+s_aa)},\\
H_2 &= \frac{U_-a}{2\pi L} \sum_{\ell, s_a=\pm 1} \i\ell \eta_{-\ell} \eta_\ell \int \normord{\powe{2\i\ell\phi(x)} \partial_x\phi(x+s_aa)},\\
H_3 &= \frac{U_+a}{2\pi^2} \sum_{\ell, s_a=\pm 1} \int \frac{1}{4(a^2+\alpha^2)}\normord{\powe{2\i\ell\klb*{\phi(x) - \phi(x+s_aa)}}},\\
H_4 &= \frac{U_+a}{2\pi^2} \sum_{\ell, s_a=\pm 1} \int \frac{4\pi^4(a^2+\alpha^2)}{L^4} \normord{\powe{2\i\ell\klb*{\phi(x) + \phi(x+s_aa)}}}.
\end{align*}
At first glance taking the continuum limit seems to be an easy task: keep $u_\pm = U_\pm a$ fixed while $a\rightarrow 0$.
There are, however, several issues with this procedure.
The asymptotic expansion of $H_3$ boils down to
\begin{align*}
&\normord{\powe{2\i\ell \klb*{\phi(x) - \phi(x+a)}}}\\
&\qquad \asymeq{a\rightarrow 0} \normord{\powe{-2\i\ell a\partial_x\phi - 2\i\ell \frac{a^2}{2} \partial_x^2\phi}} \\
&\qquad \asymeq{a\rightarrow 0} \normord{\klb*{1 - 2\i\ell a\partial_x\phi - \i\ell a^2\partial_x^2\phi - 2a^2 \kla*{\partial_x\phi}^2}} \\
&\qquad = - 2a^2 \normord{\kla*{\partial_x\phi}^2} + \text{constants and total derivatives}.
\end{align*}
Summing on $\ell$ gives $-4a^2\normord{\kla*{\partial_x\phi}^2}$, which would exactly cancel the term from the continuum limit of $H_1$ if not for the modified denominator $\frac{1}{a^2+\alpha^2}$.

Even subtleties involving $\alpha$ aside, the continuum limit for $H_2$ is much more troublesome.
The integrand turns into the total derivative $\asympropto{} \partial_x\powe{2\i\ell\phi(x)}$ and therefore the contribution vanishes.
The result would then no longer depend on the difference between the interaction strengths, but we already know this to be a crucial ingredient from the numerical analysis in \cite{EP_DynamicInduced_Quenched_Lehmann_2021}.
For now, we tentatively accept that this limit is not straight-forward and simply keep the exact expression.

Absorbing the density-density terms $(\partial_x\phi)^2$ from $H_1$ and $H_3$ into $H_0$ by modifying the coefficients then leads to the \gls{ell} presented in \autoref{sec:boson}.

\section{\acrlong{gf} of the \gls{ell}}\label{sec:app:special_func}
In \autoref{sec:app:overview_ell_gf}, we give a brief overview of the technical results that were skipped during the evaluation of the \gls{gf} of the \gls{ell} in \autoref{sec:gf}.
In \autoref{sec:app:detailed_ell_gf}, we then derive these results in more detail.
\subsection{Overview of the results}\label{sec:app:overview_ell_gf}
Computing correlators can then be done directly on the operator level and we define $\tilde{\Phi}(y) = \bra*{\varphi_\R(y) \varphi_\R(0)}_0$.
Dropping the label $\ell=\R$ for notational convenience, this evaluates to
\begin{align*}
\sum_{q,q'>0} &\frac{1}{\sqrt{n_qn_{q'}}} \powe{-\alpha \frac{q+q'}{2} } \bra*{\kla*{\powe{-qy} b_q + \powe{qy} b_q^\dagger} \kla*{b_{q'} + b_{q'}^\dagger}}_0 \\
&= \sum_{q>0} \frac{1}{n_q} \powe{-\alpha q} \klb*{\frac{2\cosh(qy)}{\powe{\beta vq} - 1} + \powe{-qy}}.
\end{align*}
where $y=v\tau+\i x$.
Expanding the hyperbolic cosine and the Bose function yields \footnote{$\sum_{n\ge 1} \frac{1}{n} x^n = -\log\kla*{1 - x}$ for $|x|<1$}
\begin{align*}
\sum_{n_q\ge 1} &\frac{1}{n_q} \powe{- \frac{2\pi}{L} n_q(\alpha + y)} + \sum_{s=\pm 1} \sum_{m\ge 1} \sum_{n_q\ge 1} \frac{1}{n_q} \powe{-\frac{2\pi}{L} n_q (\alpha + \beta vm + sy)} \\
&= \bsplit{&-\log\klb*{1 - \powe{-\frac{2\pi}{L}(\alpha + y)}} \\
&- \sum_{s=\pm 1} \sum_{m\ge 1} \log\klb*{1 - \powe{-\frac{2\pi}{L} (\alpha + \beta vm + sy)}}.}
\end{align*}
In the thermodynamic limit, this result is plagued by an \gls{ir} divergence, so we instead consider the regularized expression $\Phi(z) = \tilde{\Phi}(z) - \tilde{\Phi}(0)$,  which gives
\begin{align*}
-\log\klb*{1 + \frac{z}{w}} - \sum_{m\ge 1} \log \klb*{1 - \kla[\Big]{\frac{z}{w + \pi m}}^2}
\end{align*}
where $z=\frac{\pi}{\beta v} y$ ($z_+$ in \autoref{sec:gf}).
Utilizing the product representation of the Gamma function one can readily verify that $\prod_{m\ge 1} \kla*{1 - \frac{x^2}{(a+m)^2}} = \frac{\Gamma(1+a)^2}{\Gamma(1+a+x)\Gamma(1+a-x)}$, which yields the expression stated in \autoref{sec:gf}.

For the calculation of the single-particle \gls{gf} of the \gls{ell}, we define the set of special functions
\begin{align*}
J_{b,N}(k,\omega) &= \powe{-\i \frac{\pi}{2}N} \sin(\pi b) I_b\kla*{\omega+k} I_{b+N}\kla*{\omega-k} \numberthis \label{eqn:app:J_bN_function},
\end{align*}
where $I_b(z) = 2^{b-1}\Gamma(1-b) \frac{\Gamma\kla*{\frac{b}{2} - \i \frac{z}{4}}}{\Gamma\kla*{1 - \frac{b}{2} - \i \frac{z}{4}}}$.
One can show that at $\omega=\i 2n$ (for $N$ even) and at $\omega=\i (2n+1)$ (for $N$ odd), these functions admit the integral representation (see \appref{sec:app:detailed_ell_gf})
\begin{align*}
\intx{0}{\pi}{\intr{\powe{\omega \tau - \i k x} \csc\kla*{z_+}^{b+N} \csc\kla*{z_-}^b }{x} }{\tau},
\end{align*}
where $z_\pm = \tau \pm \i x$ and $n\in\mathbb{Z}$.
Thus, $J_{b,1}$ corresponds to the \gls{ft} of the free \gls{gf}.
At first order \acrlong{pt}, we encounter the integral
\begin{align*}
\sum_{s_a=\pm 1} \int_{0}^{\pi}\int_\mathbb{R} \powe{\omega\tau - \i kx} \csc(z_+)^{b+1} \csc(z_-)^b \cot(z_s + \i s_aa)
\end{align*}
with $\omega=\i (2n+1)$.
The solution is given by (see \appref{sec:app:detailed_ell_gf})
\begin{align*}
S_{b,1,s}(k,\omega,a)&=\frac{2}{\sinh(a)} \sum_{m\in\mathbb{Z}} \klc[\bigg]{P_{1} + \sum_{l\ge 0} \sum_{\ell=\pm 1} C_\ell P_{2}},
\end{align*}
where, with $\delta\equiv \delta_{s,-1}$ and $k_l = k + 2\i\ell (2l + m + b - \ell\delta)$,
\begin{align*}
P_{1} &= \bsplit{&J_{b-\delta, 2\delta}(k -2\i ms + s\omega, 2\i m) \\
&\quad\times \powe{-\i (\omega + 2m)a \sgn(2m - \Im \omega)},}\\
P_{2} &=  \frac{\powe{\i k_l a \sgn(\Im k_l)}}{k_l + s \omega - 2\i ms},\\
C_\ell &= \bsplit{&-\i \sin(\pi b) 4^b \frac{\Gamma(1-b+\delta)}{\Gamma(b+\delta)} \\
&\quad\times \frac{\Gamma(b+l+m)}{\Gamma(1-\ell\delta + l + m)} \frac{\Gamma(b-\ell\delta + l)}{l!}.}
\end{align*}
\subsection{Detailed derivations}\label{sec:app:detailed_ell_gf}
\paragraph*{Derivation of $J_{b,N}(k,\omega)$.}
In the following, one should think of the function $J_{b,N}$ as being defined by the integral representation; the goal is to show that this reduces to \autoref{eqn:app:J_bN_function}.
It is sufficient to solve the integral for a single $N$, because one can then recursively obtain the general result.
To illustrate the idea behind this recursion, consider $N$ to be odd and note that $J_{b,N-1}(k,\i 2n)$ equals
\begin{align*}
&\intx{0}{\pi}{\intr{\powe{\i 2n\tau - \i kx} \csc(z_+)^{b+N-1} \csc(z_-)^b}{x} }{\tau}\\
& \quad = \int_{0}{\pi} \int_{\mathbb{R}} \powe{\i 2n\tau - \i kx} \csc(z_+)^{b+N} \csc(z_-)^b \frac{\powe{\i \tau - x} - \powe{-\i\tau + x}}{2\i}\\
& \quad = \frac{1}{2\i} J_{b,N}(k-\i, \i 2n + \i) - \frac{1}{2\i} J_{b,N}(k+\i, \i 2n - \i).
\end{align*}
Of course a similar relation holds for $N$ even, as well as different frequency arguments.
In the spirit of an inductive argument, we assume that \autoref{eqn:app:J_bN_function} holds for $N$ and insert it to get
\begin{align*}
J_{b,N-1}(k,\i 2n) &= \bsplit{&\pi \powe{\i \frac{\pi}{2}(N-1)} 2^{b-1+(N-1)} \frac{I_b\kla*{\i 2n + k}}{\Gamma(b+N)} \\
&\times\klc*{f_+\kla*{\frac{\i 2n - k}{4}} - f_-\kla*{\frac{\i 2n - k}{4}} },}
\end{align*}
where $f_\pm(z) = \Gamma\kla*{\frac{b+N}{2} - \i z \pm  \frac{1}{2}} / \Gamma\kla*{1-\frac{b+N}{2} - \i z \pm \frac{1}{2}}$.
Using the recursive property of the Gamma function, $\Gamma(z+1)=z\Gamma(z)$, these can be rewritten as
\begin{align*}
f_+(z) &= \frac{\Gamma\kla*{\frac{b+N-1}{2} - \i z + 1} }{\Gamma\kla*{1-\frac{b+N-1}{2} - \i z}} \\
&= \frac{\Gamma\kla*{\frac{b+N-1}{2} - \i z} }{\Gamma\kla*{1-\frac{b+N-1}{2} - \i z}} \kla*{\frac{b+N-1}{2} - \i z},\\
f_-(z) &= \frac{\Gamma\kla*{\frac{b+N-1}{2} - \i z} }{\Gamma\kla*{1-\frac{b+N-1}{2} - \i z - 1}} \\
&= \frac{\Gamma\kla*{\frac{b+N-1}{2} - \i z} }{\Gamma\kla*{1-\frac{b+N-1}{2} - \i z}} \kla*{-\frac{b+N-1}{2} - \i z}.
\end{align*}
Plugging this into our previous result yields
\begin{align*}
J_{b,N-1}(k,\i 2n) &= \bsplit{&\pi \powe{\i \frac{\pi}{2}(N-1)} 2^{b-1+(N-1)} \frac{b+N-1}{\Gamma(b+N)} \\
&\times I_b\kla*{\i 2n + k} \frac{\Gamma\kla*{\frac{b+N-1}{2} - \i \frac{\i 2n - k}{4}} }{\Gamma\kla*{1-\frac{b+N-1}{2} - \i \frac{\i 2n - k}{4}}},}
\end{align*}
which corresponds to \autoref{eqn:app:J_bN_function}.
Similarly, one can verify that the relation also holds for $N\rightarrow N+1$, and of course the case of even $N$ is analogous too.

Now all that is left to do is to solve the integral for some $N$.
While a direct calculation for $N=0$ might be doable due to the additional symmetry of the integral, there exists a more elegant approach for the case of $N=1$.
Since this is equivalent to the \gls{ft} of \autoref{eqn:gf_order0_matsubara_real_space}, it must correspond to the retarded \gls{gf} of an interacting \gls{ll} (whose \gls{ft} turns out to be much easier to compute).
Similar to the calculation for the Matsubara \gls{gf}, we define $\xi_\pm = \frac{\pi}{\beta v} \kla*{vt \pm x}$ and find that $\bra[\big]{\psi_\ell(\xi_+, \xi_-) \psi_{\ell'}^\dagger(0,0)}_0$ yields
\begin{align*}
&\frac{2\delta_{\ell,\ell'}}{2\pi\alpha} \prod_{s=\pm 1} \powe{-\frac{1}{4} \klb*{-K - \frac{1}{K} + 2\ell s} \Phi(\xi_s)} \\
&\quad = \frac{2\delta_{\ell,\ell'}}{2\pi\alpha} \prod_{s=\pm 1} \klb*{-\i w \csch(\xi_s)}^{M+\delta_{-\ell,s}} \\
&\quad \asymeq{\alpha\rightarrow 0} \frac{\delta_{\ell,\ell'} w^{2M}}{\beta v\i} \powe{-\i\pi M} \csch(\xi_{-\ell})^{M+1} \csch(\xi_{\ell})^M.
\end{align*}
The retarded \gls{gf} also contains a similar term from the anti-commutator, which we can obtain from the above by substituting $\xi_\pm \rightarrow - \xi_\pm$.
This replacement has a rather subtle consequence that originates from the fractional exponentiation of real numbers:
In our case, this leads to the phase factors
\begin{align*}
&\sinh(-\xi_{-\ell})^{M+1} \sinh(-\xi_{\ell})^M \\
&\quad = \powe{\i\pi (M+1) \sgn(\xi_{-\ell})} \sinh(\xi_{-\ell})^{M+1} \powe{\i\pi M \sgn(\xi_{\ell})} \sinh(\xi_{\ell})^M,
\end{align*}
and hence a global minus sign if $\xi_+\xi_- < 0$.
Combining the two terms is thus equivalent to multiplication by an additional factor of the form
\begin{align*}
&1 + \powe{\i\pi (M+1) \sgn(\xi_{-\ell}) + \i\pi M \sgn(\xi_{\ell})} \\
&\quad = \Theta(\xi_+\xi_-) \klc*{1 - \powe{\i 2\pi M \sgn(\xi_{-\ell}) }} = -\powe{\i \pi M} 2\i\sin(\pi M)
\end{align*}
for $\xi_\ell\ge 0$.
The latter is a valid assumption, because the retarded \gls{gf} contains a factor of $\Theta(t) = \Theta(\xi_++\xi_-)$, and thus $G^\ret_{0,\ell,\ell'}(\xi_+, \xi_-) = -\i\Theta(t) \bra[\big]{\klc[\big]{\psi_\ell(\xi_+, \xi_-), \psi_{\ell'}^\dagger(0)}}_0$ evaluates to (only $\ell=\ell'$ is non-zero)
\begin{align*}
\i\Theta(t) \Theta(\xi_+\xi_-) \frac{w^{2M}}{\beta v} 2\sin(\pi M) \csch(\xi_{-\ell})^{M+1} \csch(\xi_{\ell})^M,
\end{align*}
where $\Theta(t) \equiv \Theta(\xi_+ + \xi_-)$.
The two-dimensional \gls{ft} can also be rewritten in terms of the light-cone variables $\xi_\pm$, i.e. $G^\ret_{0,\ell,\ell}(k,\omega) = \intr{ \intr{\powe{\i \omega t - \i kx} G^\ret_{0,\ell,\ell}(x,t)}{x} }{t}$ becomes
\begin{align*}
\frac{\beta^2}{2v\pi^2} \int_{0}^{\infty} \int_{0}^{\infty} \powe{\i\omega \frac{\beta}{2\pi} (\xi_+ + \xi_-) - \i k \frac{\beta v}{2\pi} (\xi_+ - \xi_-)} G^\ret_{0,\ell,\ell}(\xi_+, \xi_-) .
\end{align*}
This is valid, because the Heaviside function $\Theta(t)$ limits the integration region to $t\ge 0$, which corresponds to $\xi_\pm \ge 0$.
With $k_\pm = \frac{\beta}{\pi} (\omega \pm vk)$, inserting our expression for the retarded \gls{gf} and using $\intx{0}{\infty}{\powe{\i zx/2} \csch(x)^b }{x} = I_b(z)$ finally yields
\begin{align*}
G^\ret_{0,\ell,\ell}(k,\omega) &= \frac{\i\beta^2}{2v\pi^2} \frac{w^{2M}}{\beta v} 2\sin(\pi M) I_M(k_{-\ell}) I_{M+1}(k_{\ell}).
\end{align*}
The idea at this point is to realize that, if the analytic continuation $\i\omega_n\rightarrow \omega + \i \eta$ is possible for the result of the \gls{ft} of \autoref{eqn:gf_order0_matsubara_real_space}, it must be identical to the above.
More explicitly, rescaling the integration variables gives
\begin{align*}
&G^\M_{0,\ell,\ell}(k,\i\omega_n^\F) = \intx{0}{\beta}{\intr{\powe{\i\omega_n^\F\tau - \i kx} G^\M_{0,\ell,\ell}(x,\tau)}{x} }{\tau} = \\
&\frac{-w^{2M}}{2\beta v} \frac{\beta^2}{v\pi^2} \int_{0}^{\pi}\int_{\mathbb{R}} \powe{\i \omega_n^\F \frac{\beta }{\pi} \tau + \i k \ell \frac{\beta v}{\pi} x} \csc\kla*{z_+}^{M+1} \csc\kla*{z_-}^M.
\end{align*}
This is consistent with our result for the retarded \gls{gf} if, and only if,
\begin{align*}
&\intx{0}{\pi}{\intr{\powe{\i (2n+1) \tau - \i k x} \csc\kla*{\tau + \i x}^{M+1} \csc\kla*{\tau - \i x}^M }{x} }{\tau} \\
&\quad = -\i \sin(\pi M) I_M\kla*{\i(2n+1) + k} I_{M+1}\kla*{\i(2n+1) - k},
\end{align*}
which is clearly equivalent to \autoref{eqn:app:J_bN_function} for $N=1$.
\paragraph*{Derivation of $S_{b,1,s}(k,\omega,a)$.}
The integral representation only holds $\omega = \i w_n$, where $w_n = 2n+1$.
Since we already know the \gls{ft} without the extra cotangent, it would be natural to split off this factor via the \gls{ct}.
We choose to split off an additional cosecant factor as well, because this later makes the convolution in momentum space more symmetric.
This is made possible by the remarkable identity
\begin{align*}
T &= \sum_{s_a=\pm 1} \intx{0}{\pi}{\intr{\powe{\i w_n \tau - \i kx} \csc(z_+) \cot(z_+ + \i s_aa)}{x} }{\tau}\\
&= \frac{4\pi}{\i w_n + k} \frac{\sin(ka)}{\sinh(a)}.
\end{align*}
We only consider the case $n\ge 0$ here.
Reading the complex substitution $z=\powe{\i 2\tau}$ backwards, the $\tau$-integration turns into a contour integral around the unit disk,
\begin{align*}
T&= \sum_{s_a=\pm 1} \oint_{|z|=1} \intr{ z^n\powe{- \i kx} \frac{\powe{x}}{z - \powe{2x}} \i \frac{z + \powe{2(x+s_aa)}}{z - \powe{2(x+s_aa)}} }{x} \,\d z.
\end{align*}
The contribution from the first pole at $z=\powe{2x}$ vanishes upon summation over $s_a$, since
\begin{align*}
\sum_{s_a=\pm 1} &2\pi\i \intx{-\infty}{0}{\powe{(w_n-\i k)x} \i \frac{1 + \powe{2s_aa}}{1 - \powe{2s_aa}}}{x} \\
&= \sum_{s_a=\pm 1} \frac{2\pi \coth(s_aa)}{w_n -\i k} = 0.
\end{align*}
For the second pole at $z=\powe{2(x+s_aa)}$, we instead obtain
\begin{align*}
\sum_{s_a=\pm 1} &2\pi\i \intx{-\infty}{-s_aa}{\powe{(w_n-\i k)x} \i\csch(s_aa) \powe{s_aaw_n}}{x} \\
&= \frac{2\pi}{\i w_n + k} \sum (-\i) \frac{s_a\powe{s_aaw_n}}{\sinh(a)} \powe{-s_aa(w_n-\i k)} \\
&= \frac{4\pi}{\i w_n + k} \frac{\sin(ka)}{\sinh(a)}.
\end{align*}
Combined with the \gls{ft} of the remaining cosecant-product and $\delta \equiv \delta_{s,-1}$, the \gls{ct} now gives
\begin{align*}
S_{b,1,s} &= \frac{1}{2\pi^2} \sum_{m\in\mathbb{Z}} \bsplit{\intr{ &J_{b - \delta, 2\delta}(k-k', 2\i m)\\
&\times \frac{4\pi}{s\i w_{n-m} + k'} \frac{\sin(k'a)}{\sinh(a)}}{k'}.}
\end{align*}
The integrand only contains simple poles, albeit infinitely many.
Thus a cumbersome, but conceptually straight-forward solution via the residue theorem is possible.
Alternatively, one could perform the Matsubara sum first, but the result seems to perform quite a bit worse with respect to the rate of convergence.

For the oscillatory factor $\powe{\i k'a}$, the contour has to be closed in the upper half of the complex plane to ensure that the integral along the arc vanishes as the radius tends to infinity.
Similarly, for $\powe{-\i k'a}$ we have to close it in the lower half and include an extra minus sign due to the changed orientation.
As before, we consider the poles separately, starting with the simple one at $k'=-s\i w_{n-m}$.
Its residue is
\begin{align*}
\frac{1}{2\i} \powe{sw_{n-m} a} J_{b-\delta, 2\delta}(k+s\i w_{n-m}, 2\i m) \Theta\kla*{- s\Im \i w_{n-m}} 
\end{align*}
for $\powe{\i k'a}$, and for $\powe{-\i k'a}$ it is
\begin{align*}
\frac{-1}{2\i} \powe{-sw_{n-m} a} J_{b-\delta, 2\delta}(k+s\i w_{n-m}, 2\i m) \Theta\kla*{s\Im \i w_{n-m}} .
\end{align*}
Combining both contributions then yields the first part of our result,
\begin{align*}
\frac{2}{\sinh(a)} \sum_{m\in\mathbb{Z}} \bsplit{&J_{b-\delta, 2\delta}(k+s\i w_{n-m}, 2\i m) \\
&\times \powe{w_{n-m} a \sgn(-\Im \i w_{n-m})}.}
\end{align*}
In order to deal with the other poles, we start by writing
\begin{align*}
J_{b,N}(k,\omega) &= C_{b,N} \prod_{\ell=\pm 1} \frac{\Gamma\kla[\big]{\frac{b_\ell}{2} - \i k_\ell}}{\Gamma\kla[\big]{1 - \frac{b_\ell}{2} - \i k_\ell}},
\end{align*}
where $k_\ell = \frac{\omega + \ell k}{4}$, $C_{b,N} = \pi \i^N 2^{2b-2+N} \frac{\Gamma(1-b)}{\Gamma(b+N)}$ and $b_\ell = b + \frac{1 - \ell}{2}N$.
The poles correspond to the zeros of the argument of the Gamma functions in the numerator, and are thus located at $-l = \frac{b_\ell}{2} - \i k_\ell$, or equivalently,
\begin{align*}
k_l &= -2\i\ell (2l + b_\ell) - \ell\omega,
\end{align*}
where $l\ge 0$.
Using $\eta \Gamma(-l+\eta) \rightarrow \frac{(-1)^l}{l!}$, the residue of $J_{b,N}$ at these poles, i.e. $\Res[J_{b,N}](k_l)$, becomes
\begin{align*}
C_{b,N} &\lim_{\eta\rightarrow 0} \eta \frac{\Gamma\kla[\big]{\frac{b_{\ell}}{2} - \i \frac{\omega + \ell (k_l + \eta)}{4}}}{\Gamma\kla[\big]{1 - \frac{b_{\ell}}{2} - \i \frac{\omega + \ell k_l}{4}}} \frac{\Gamma\kla[\big]{\frac{b_{-\ell}}{2} - \i \frac{\omega - \ell k_l}{4}}}{\Gamma\kla[\big]{1 - \frac{b_{-\ell}}{2} - \i \frac{\omega - \ell k_l}{4}}}\\
&= C_{b,N} \frac{(-1)^l}{l!} \frac{-4\i \ell}{\Gamma\kla{1 - b_\ell - l}} \frac{\Gamma\kla[\big]{b + \frac{N}{2} + l -\i \frac{\omega}{2}}}{\Gamma\kla[\big]{1 - \ell \frac{N}{2} + l -\i \frac{\omega}{2}}}.
\end{align*}
In our case, $\omega=2\i m$, and due to the argument being $k-k'$ rather than $k'$, the poles are at $k_l = k + 2\i\ell (2l+m+b_\ell)$ with $b_\ell = b - \delta + \frac{1-\ell}{2} 2\delta = b - \delta \ell$.
The contribution from these poles is thus given by
\begin{align*}
\frac{2}{\sinh(a)} \sum_{m\in\mathbb{Z}} \sum_{l\ge 0} \sum_{\ell=\pm 1} \bsplit{&\frac{(-1)^l}{l!} \frac{-4\i\ell C_{b - \delta,2\delta} }{\Gamma(1-b + \delta_{s,-1}\ell - l)} \\
&\times \frac{\Gamma\kla*{b + l +m}}{\Gamma\kla*{1 - \delta \ell + l +m}} \frac{\powe{\i k_l a \sgn(\Im k_l)}}{k_l + s\i w_{n-m}}.}
\end{align*}
Performing the analytic continuation $\i (2n+1)\rightarrow \omega + \i \eta$ finally yields the result stated in \appref{sec:app:overview_ell_gf}.

As a final remark, the complexity of the integrand makes a closed solution unlikely, so some series representation is probably unavoidable (although there may exist various different forms).
The main difficulty lies in the shift of the cotangent argument, hence the integral is not analytic at $a=0$, and it is to be expected that small values of $a$ are harder to resolve.

\section{RG Flow Equations}\label{sec:app:rg}
Here, we provide additional details for our momentum-shell \gls{rg} and also consider one-loop corrections.
To do so, we use the prescription for an arbitrary momentum cut-off for the special case of $d=2$ \cite[\spage 690]{RG_QFT_Dupuis_2025}:
Every integral in momentum space is replaced by
\begin{align*}
\int_\textbf{k} &\longrightarrow \int_\textbf{k} f\kla*{\frac{k^2}{\Lambda^2}},
\end{align*}
where $\Lambda$ is the UV regulator \cite{QFT_RG_OPE_Collins_1984} and $f(x)$ describes the arbitrary cut-off, admitting the boundary conditions $f(0)=1$ and $f(\infty)=0$.
In the \gls{rg} step, the integral over fast modes then becomes
\begin{align*}
\int_\textbf{k} \klb[\bigg]{f\kla*{\frac{k^2}{\Lambda^2}} - f\kla[\bigg]{\frac{k^2}{\Lambda^2\powe{-2\d l}}}} = -2\d l \frac{1}{\Lambda^2} \int_\textbf{k} k^2 f'\kla*{\frac{k^2}{\Lambda^2}}.
\end{align*}
The flow equations read \footnote{Note that, if the integrals converge, $f'(0)$ is the result from $\int \tilde{\Phi}_\f(\textbf{r})\,\d^2r = -\frac{4\pi}{\Lambda^2}f'(0)$ with $\tilde{\Phi}_\f(\textbf{r}) = -\frac{1}{\pi \Lambda^2} \int \powe{\i \textbf{r}\cdot \textbf{q}} f'\big(\frac{q^2}{\Lambda^2}\big)\,\d^2q$. A sharp cut-off $f(x)=\Theta(1-x)$ instead leads to $\tilde{\Phi}_\f(\textbf{r}) = J_0(\Lambda r)$ and the indeterminate limit $\int \tilde{\Phi}_\f(\textbf{r}) \propto \lim_{R\rightarrow \infty} RJ_1(\Lambda R)$.}
\begin{align*}
\frac{\d g_4}{\d l} &= \kla*{2 - 4K} g_4,\\
\frac{\d g_2}{\d l} &= \kla*{1 - K} g_2 + 32\pi K g_4 g_2 f'(0),\\
\frac{\d v}{\d l} &= -16\pi^2 K^2 v g_2^2 \klb*{2 + C_1(K)},\\
\frac{\d K}{\d l} &= 16\pi^2 K^3 \kla*{ g_2^2 \klb*{2 + C_1(K)} - 8 g_4^2 C_3(4K)},
\end{align*}
where at \gls{rg}-time $l=0$, we have $g_4 = \frac{U_+a}{4\pi^2v}$ and $g_2 = \frac{U_-a}{4\pi^2v}$.
The functions $C_n(p)$ are given by
\begin{align*}
-\intx{0}{\infty}{f'(t) \intx{0}{\infty}{J_0(s\sqrt{t}) s^n \powe{p \intx{0}{\infty}{f(u) \frac{1}{u} \klb*{J_0(s\sqrt{u}) - 1}}{u}}}{s} }{t},
\end{align*}
where $J_0$ is a Bessel function.
This exemplifies that one-loop corrections depend on the choice of the cut-off function; for $f(x)=\powe{-x}$ from \cite{RG_QFT_Dupuis_2025}, we get $C_n(p) = 2^n \powe{-p(\gamma + \log 2)} \intx{0}{\infty}{u^{\frac{n-1}{2}-p} \powe{-pE_1(u)-u}}{u}$.
Meanwhile, our result reproduces \cite[\seqn E.23]{Giamarchi_2004} for a sharp cut-off ($f(x)=\Theta(1-x)$).
Aside from using an arbitrary cut-off, the calculation essentially follows \cite[\sapp E]{Giamarchi_2004}, so we do not reproduce it here.

While the dependence of the coefficients on the cut-off makes quantitative predictions difficult, the qualitative behavior of the parameters in \autoref{fig:ll_fpt_fit} and \autoref{fig:ll_num_fit} can at least be partially understood.
The observed increase in $K$ with decreasing $U_+$ is consistent with the \gls{rg} flow, and so is the decrease of $v$ with increasing $U_-$.
However, the exponential dependence of the effective coupling $g$ on $U_-$ is harder to reconcile, and going beyond $U_\A=0$ seems to be fundamentally out of reach of a perturbative approach.

\section{Fermionic \acrlong{pt}}\label{sec:app:fermionic_pt}
Here, we derive the \gls{gf} of the model via fermionic plain \gls{pt} to second order in the interaction.
All quantities are assumed to be in the $\A,\B$-basis, and we set the lattice spacing to $a=1$.
\subsection{Symmetries of the Model Hamiltonian}
The original microscopic model from the main text is $H=H_0 + H_\tint$.
In momentum space, the free part is $H_0 = \sum_k \textbf{c}_k^\dagger H_0 (k) 
\textbf{c}_k$, where $\textbf{c}_k = (c_{k,\A}, c_{k,\B})^\transpose$ and
\begin{align*}
H_0 (k) &= [t_x\cos(k) - J] \sigma_x + t_y \sin(k) \sigma_y
\end{align*}
and the interaction reads
\begin{align*}
H_\mathrm{int} &= \sum_{s = \A,\B} \sum_{j} U_s \kla*{n_{j, s} - \frac{1}{2}} \kla*{n_{j + 1, s} - \frac{1}{2}}.
\end{align*}
The terms $H_0$ and $H_\tint$ both exhibit \gls{trs} in the form of 
\begin{align*}
\text{\gls{trs}}\quad : \quad  c_{j,s} &\to c_{j, s}, \quad &c^\dagger_{j, s} &\to c^\dagger_{j, s}, \quad &\i &\to -\i, \\
c_{k,s} &\to c_{-k, s}, \quad &c^\dagger_{k, s} &\to c^\dagger_{-k, s}, \quad &\i &\to -\i.
\end{align*}
The mapping in the last column means complex conjugation. 
Here, invariance under \gls{trs} is tantamount to the Hamiltonian being expressible as a sum over strings of real space operators with real coefficients. 
The second line is the action of \gls{trs} on the $k$-space operators $c_{k, s}^\dagger = \frac{1}{\sqrt{N}} \sum_j e^{\i kj} c_{j, s}^\dagger$.

Both $H_0$ and $H_\tint$ also possess the \gls{phs} described by
\begin{align*}
\text{\gls{phs}}\quad : \quad  c_{j, \A} &\to c_{j, \A}^\dagger, \quad &c_{j, \B} & \to -c_{j, \B}^\dagger,\\
c_{k, \A} &\to c_{-k, \A}^\dagger, & \qquad c_{k, \B} &\to -c^\dagger_{-k, \B}.
\end{align*}
The combination of \gls{trs} and \gls{phs} implies the \gls{cs},
\begin{align*}
\text{\gls{cs}} \quad : \quad c_{j, \A} &\to c_{j, \A}^\dagger, \quad &c_{j, \B} & \to -c_{j, \B}^\dagger, \quad &\i &\to -\i, \\
c_{k, \A} &\to c_{k, \A}^\dagger, \quad &c_{k, \B} &\to -c^\dagger_{k, \B}, \quad &\i &\to -\i.
\end{align*}
The consequences of these symmetries for the \gls{gf} at general complex $z$ are (see \appref{sec:app:symmetries_kl} for a derivation)
\begin{align*}
\text{\gls{trs}} \quad&:\quad & G(k, z)^\transpose &= G(-k, z), \\
\text{\gls{phs}} \quad&:\quad & G(k, z)^\transpose &= -\sigma_z G(-k, -z) \sigma_z, \\
\text{\gls{cs}} \quad&:\quad & G(k, z) &= -\sigma_z \left(G(k, -z) \right) \sigma_z,
\end{align*}
The symmetries for the \gls{gf} directly carry over to the effective Hamiltonian $H_\eff(k, z) = z - G(k, z)^{-1}$ by elementary algebraic manipulation.

We note that a perturbative symmetry-respecting expansion of the the self-energy in powers of $U_s$,
\begin{align*}
\Sigma(k, z) &= U_s \Sigma^{(1),s}(k, z) + U_s U_{s'} \Sigma^{(2), s, s'}(k, z) + \dots,
\end{align*}
must obey these symmetries order by order, i.e. already on the level of the individual $2\times 2$-matrices $\Sigma^{(n), s_1,\dots, s_n}(k, z)$.

\subsection{\acrlong{pt} for Pair Interactions}
To apply standard fermionic \gls{pt} for a density-density interaction (see e.g. \cite{QMB_Bruus_2004}), we shift the chemical potential terms arising on each sublattice from the PH-symmetric interaction into the quadratic part.
This yields $H_0'=\sum_k \textbf{c}_k^\dagger H_0'(k) 
\textbf{c}_k$, where
\begin{align*}
H_0'(k) &= [t_x\cos(k) - J] \sigma_x + t_y \sin(k) \sigma_y - \begin{psmallmatrix}U_\A&0\\0&U_\B\end{psmallmatrix},
\end{align*}
and the remaining density-density term takes the form
\begin{align*}
H_\tint' &= \sum_{s = \A,\B} \sum_{j} U_s n_{j, s} n_{j + 1, s}  \\
&= \frac{1}{2} \sum_{s,s'} \sum_{j,j'} U_{s, s'}(j-j') n_{j, s} n_{j', s}, \numberthis \label{eqn:app:U_int}
\end{align*}
where $U_{s, s'}(j-j') = \delta_{s, s'} U_s (\delta_{j, j' + 1} + \delta_{j+1, j'})$.
The resulting self-energy correction to $H_0'$, i.e. $\Sigma'_{s, s'}(k, \i \omega_n)$, can be schematically represented as the sum over all topologically inequivalent Feynman diagrams,
\begin{align}
\SHTnoLabel{0.12}  + \SFnoLabel{0.12} + \SSoOneNoLabel{0.12} +  \SSoTwoNoLabel{0.12}+ \dots, \label{eqn:app:S_expansion_no_sym}
\end{align}
where all labels are suppressed, but summation over internal indices, Matsubara frequencies and momenta is implied.
The free propagator is given by
\begin{align*}
\GfreeU{0.15} = {G'}^0_{s, s'}(k, \i \omega_n) = \left( [\i \omega_n - H_0'(k)]^{-1} \right)_{s, s'} ,
\end{align*}
and the interaction vertex is the \gls{ft} of the interaction potential $U_{s, s'}(j-j')$ (cf. \autoref{eqn:app:U_int}),
\begin{align*}
\Vint{0.15} = \delta_{s, s'} U_s 2\cos(k)
\end{align*}
To obtain the expansion from \autoref{eqn:app:S_expansion_no_sym}, we divided the system into $H_0'$ and $H_\tint'$ which do not obey \gls{phs} on their own, so this symmetry is also not conserved at each order of the perturbative expansion.
To remedy this, we symmetrize the resulting effective Hamiltonian $H_\eff'(k, \i \omega_n) = H_0'(k) + \Sigma'(k, \i \omega_n)$ with its \gls{cs}-conjugate (since \gls{trs} is still conserved order by order, this is equivalent to symmetrizing w.r.t. \gls{phs}).
Using the relation $G(k, z) = -\sigma_z G(k, -z)  \sigma_z = -\sigma_z G(k, -z^\ast)^\dagger \sigma_z$, this amounts to setting
\begin{align*}
H_\eff(k, \i \omega_n) &= \frac{H_\eff'(k, \i \omega_n) - \sigma_z H_\eff'\kla*{k,- (\i \omega_n)^\ast}^\dagger \sigma_z}{2} \\
&=: H_0(k) + \Sigma(\i \omega_n). \numberthis \label{eqn:app:He_sym}
\end{align*}
Note that the analytical continuation of the second term to the real axis should be $- (\i \omega_n)^\ast \to - \omega + \i \eta$ to obtain the symmetrized retarded \gls{gf} at real $\omega$.

The effect of the symmetrization on \autoref{eqn:app:He_sym} is to cancel the matrix $U = - \begin{psmallmatrix}U_\A&0\\0&U_\B\end{psmallmatrix}$ from $H_0'(k)$ to turn it into $H_0(k)$, and to cancel some diagrams from \autoref{eqn:app:S_expansion_no_sym}.
In particular, this affects the diagrams that are diagonal in $\A,\B$ and do not depend on the external frequency $\i \omega_n$, which means all diagrams with an uninterrupted interaction line connecting the external vertices. 
To second order, this only leaves us with contributions of the two diagrams
\begin{align*}
\SSoOneNoLabel{0.12} + \SSoTwoNoLabel{0.12} 
\end{align*}
to the symmetrized self-energy $\Sigma(\i \omega_n)$. 

To arrange the result in powers of $U$, we must account for the $U$-dependence of the free propagator $G'^0(k, \i \omega_n)$ in form of the matrix $U$.
To obtain an expression in powers of $U_s$, we consider the propagator $G^0(k, \i \omega_n) = \klb*{\i \omega_n - H_0(k)}^{-1}$ associated with the original free Hamiltonian $H_0(k)$ and note that $G'^0(k, \i \omega_n)$ obeys the Dyson equation
\begin{align*}
G'^0 &= G^0 + G^0 U G'^0 = G^0 \sum_{l\ge 0} \kla*{U G^0}^l.
\end{align*}
The second equality follows from self insertion of the Dyson equation.
Crucially, since $H_0(k = 0) = 0$, this geometric series expansion can only be expected to hold for general $k$ if $U_s < \omega_n = \frac{(2n + 1)\pi}{\beta}$, which limits the perturbative approach to high temperatures. 

Let us introduce the symbol for the free propagator associated to $H_0(k)$,
\begin{align*}
\Gfree{0.15} = G^0_{s, s'}(k, \i \omega_n) = \left( \klb*{\i \omega_n - H_0(k)}^{-1} \right)_{s, s'}.
\end{align*}
To collect terms up to order $U^2$, we can simply use $G^0(k, \i \omega_n)$ in the two diagrams contributing to the symmetrized self-energy; any higher powers in the expansion of $G'^0(k, \i \omega_n)$ should be combined with the contributions from higher order diagrams to correctly capture higher orders of $U$. 
Thus, we need to evaluate
\begin{align*}
\SSoOneLabelled{0.4} + \SSoTwoLabelled{0.4}.
\end{align*}
This boils down to $U_s U_{s'}  D_{s, s'} (k, \i \omega_n)$, where $D_{s, s'} (k, \i \omega_n)$ denotes the $U$-independent part of the expression,
\begin{align*}
D_{s,s'} &=  \bsplit{4 &\int_{-\pi}^\pi \frac{\d q_1}{2 \pi} \int_{-\pi}^\pi \frac{\d q_2}{2 \pi} M_{s,s'}(k,\i\omega_n, q_1,q_2) \\
&\times \klb*{\cos(q_1 - q_2) \cos(k - q_1) - \cos(k - q_1)^2},}
\end{align*}
(the minus sign comes from the fermionic loop in the second diagram)  with the Matsubara summations
\begin{align*}
M_{s,s'} &= \bsplit{\frac{1}{\beta^2} \sum_{\i \nu_n, \i \nu_n'} & G^0_{s,s'}(q_1, \i \nu_n) G^0_{s', s}(q_2, \i \nu_n') \\
&\times G^0_{s, s'}(k - q_1 + q_2, \i \omega_n - \i \nu_n + \i \nu_n').}
\end{align*}
The free propagator $G_0(k, \i \omega_n)$ can be written as
\begin{align*}
G^0(k, \i \omega_n) = \frac{\i \omega_n + d_x(k) \sigma_x + d_y(k) \sigma_y}{(\i \omega_n)^2 - d^2_x(k) -d^2_y(k)^2},
\end{align*}
where $d_x(k) = t_x\cos(k) - J$, $d_y(k) =  t_y \sin(k)$ denote the $\sigma_x, \sigma_y$ coefficients appearing in $H_0(k)$.
With the dispersion $\epsilon(k) = \sqrt{d^2_x(k) + d^2_y(k)}$, the components become
\begin{align*}
G^0_{\A,\A} &= \frac{1}{2} \klc*{\frac{1}{\i \omega_n + \epsilon(k) } + \frac{1}{\i \omega_n - \epsilon(k)}} = G^0_{\B, \B}, \\
G^0_{\A,\B} &= \frac{f(k)}{2} \klc*{\frac{1}{\i \omega_n - \epsilon(k)} - \frac{1}{\i \omega_n + \epsilon(k)}} = \kla*{G^0_{\B,\A}}^\ast,
\end{align*}
where $f(k) = \frac{d_x(k) - \i d_y(k)}{\epsilon(k)}$. 

After inserting the \gls{gf} components into $M_{s,s'}$, the elementary Matsubara summation appearing in all perturbative expressions is $M_0(z_1, z_2, z_3, \i \omega_n)$, which we can rewrite as
\begin{align*}
\frac{1}{\beta^2} &\sum_{\i \nu_n, \i \nu_n'}   \frac{1}{ \i  \nu_n + z_1} \frac{1}{ \i  \nu_n' + z_2}  \frac{1}{ \i  \omega_n -  \i  \nu_n +  \i  \nu_n' + z_3} \\
&= -\frac{1}{4} \frac{1 - t_1t_2 + t_1t_3 - t_2t_3}{ \i  \omega_n + z_3 - z_2 + z_1},
\end{align*}
where $t_a \equiv \tanh(\frac{\beta z_a}{2})$ for $a=1,2,3$.
The above result follows from the Matsubara summation identity 
\begin{align*}
\frac{1}{\beta}  \sum_{\nu_n} g( \i  \nu_n) &= - \frac{1}{2} \sum_{a} \tanh \left (\frac{\beta z_a}{2} \right) \Res[g](z_a)
\end{align*}
and the relations ($n \in \mathbb{Z}$)
\begin{align*}
&\tanh(z + n  \i  \pi) = \tanh(z) = \coth\kla*{z - \frac{(2 n + 1)  \i  \pi}{2}}, \\
&\coth(x \pm y)\klb*{\tanh(x) \pm \tanh(y)} = 1 \pm \tanh(x) \tanh(y)
\end{align*}
for hyperbolic functions.
The diagonal components of $M_{s,s'}$ can be expressed through $M_0$ as
\begin{align*}
&M_{\A, \A}(k ,  \i  \omega_n) = M_{\B, \B}(k , \i \omega_n)  \\
&\quad =\frac{1}{8} \sum_{\substack{r_1, r_2, \\  r_3 = \pm 1}} M_0\kla[\big]{r_1 \epsilon(q_1),  r_2 \epsilon(q_2), r_3 \epsilon(k - q_1 + q_2),  \i  \omega_n}
\end{align*}
and the off-diagonals as
\begin{align*}
&M_{\A, \B}(k, \i \omega_n) = M_{\B, \A}(k, -\i  \omega_n)^\ast\\
&\quad = \bsplit{-\frac{1}{8} &\sum_{\substack{r_1, r_2,\\ r_3 = \pm 1}} r_1 r_2 r_3 f(q_1) f^\ast (q_2) f(k - q_1 + q_2) \\
&\ \ \times M_0\kla[\big]{r_1 \epsilon(q_1),  r_2 \epsilon(q_2), r_3 \epsilon(k - q_1 + q_2),  \i  \omega_n}.}
\end{align*}

Finally, the retarded self-energy correction at frequency $\omega$ is found by adding the CS conjugate as indicated by \autoref{eqn:app:He_sym} and performing the analytic continuation $\i \omega_n \to \omega + \i \eta$, yielding
\begin{align}
\Sigma^R(k, \omega) = 
 \begin{pmatrix}
U_\A U_\A D_\text{diag} & U_\A U_\B D_\text{offdiag} \\ U_\B U_\A D^\ast_\text{offdiag}& U_\B U_\B D_\text{diag}
\end{pmatrix} . \label{eqn:app:Sigma_ret}
\end{align}
Here, the symmetrized components of the ``skeleton'' of the self-energy are $D_\text{diag}(k, \omega) = \frac{1}{2} \klb[\big]{D_{\A,\A} (k, \omega +  \i \eta) - D^\ast_{\A,\A} (k, -\omega +  \i \eta)}$, which is given by
\begin{align*}
& \int_{-\pi}^\pi \frac{\d q_1}{2\pi} \int_{-\pi}^\pi \frac{\d q_2}{2\pi} \klb*{\cos(q_1 - q_2) - \cos(k - q_1)} \cos(k - q_1) \\
&\times \frac{1}{8} \sum_{\substack{r_1, r_2, \\  r_3 = \pm 1}} \klb*{1 - t_1 t_2 + t_1t_3 - t_2t_3}\\
& \qquad\qquad\times \frac{\omega +  \i \eta}{(\omega +  \i \eta)^2 - [r_1 \epsilon(q_1)  - r_2 \epsilon(q_2) + r_3 \epsilon(q_3)]^2 },
\end{align*}
and $D_\text{offdiag}(k, \omega) = \frac{1}{2} \klb[\big]{D_{\A,\B} (k, \omega +  \i \eta) + D^\ast_{\B,\A} (k, -\omega +  \i \eta)}$, given by
\begin{align*}
& \int_{-\pi}^\pi \frac{\d q_1}{2\pi} \int_{-\pi}^\pi \frac{\d q_2}{2\pi} \klb*{\cos(q_1 - q_2) - \cos(k - q_1)} \cos(k - q_1) \\
&\times \frac{1}{8} \sum_{\substack{r_1, r_2, \\  r_3 = \pm 1}} r_1 r_2 r_3 f(q_1) f^\ast (q_2) f(q_3) \klb*{1 - t_1 t_2 + t_1t_3 - t_2t_3}\\
& \qquad\qquad\times   \frac{r_1 \epsilon(q_1)  - r_2 \epsilon(q_2) + r_3 \epsilon(q_3)}{[r_1 \epsilon(q_1)  - r_2 \epsilon(q_2) + r_3 \epsilon(q_3)]^2 - (\omega +  \i \eta)^2},
\end{align*}
where $t_a \equiv \tanh\kla*{\frac{\beta r_a}{2} \epsilon(q_a)}$ and $q_3 = k-q_1+q_2$.
The above results follow immediately from the previous definitions and the identities 
\begin{align*}
\frac{1}{\omega + \i \eta + z} - \frac{1}{-\omega - \i \eta + z} &= 2 \frac{\omega + \i \eta}{(\omega + \i \eta)^2 - z^2},\\
\frac{1}{\omega + \i \eta + z} + \frac{1}{-\omega - \i \eta + z} &= 2 \frac{z}{z^2 - (\omega + \i \eta)^2}.
\end{align*}

\subsection{Comparison to Data from \acrlong{csba}}
Here, we compare the results of the analytical approach as per \autoref{eqn:app:Sigma_ret} to the numerical data from \gls{csba}. 

Note that the \gls{csba} result is obtained by calculating the retarded self-energy on the real time domain and Fourier transforming the result as $\Sigma^\text{\gls{csba}} (k, \omega) = \intx{0}{\infty}{\powe{\i \omega t} \Sigma^\text{\ret, \gls{csba}}(k, t)}{t}$.
Since \gls{csba} captures Feynman diagrams beyond plain second order \gls{pt}, including higher order decay processes, we expect an excitation to equilibrate faster in \gls{csba}. Further, there is the well-known artificial, $k$-dependent damping \cite{Friesen2010} within the \gls{csba} real-time formalism.
This essentially corresponds to the finite $\eta$ that is used to regularize the analytic result.

\begin{figure}[htp]
\centering
\includegraphics{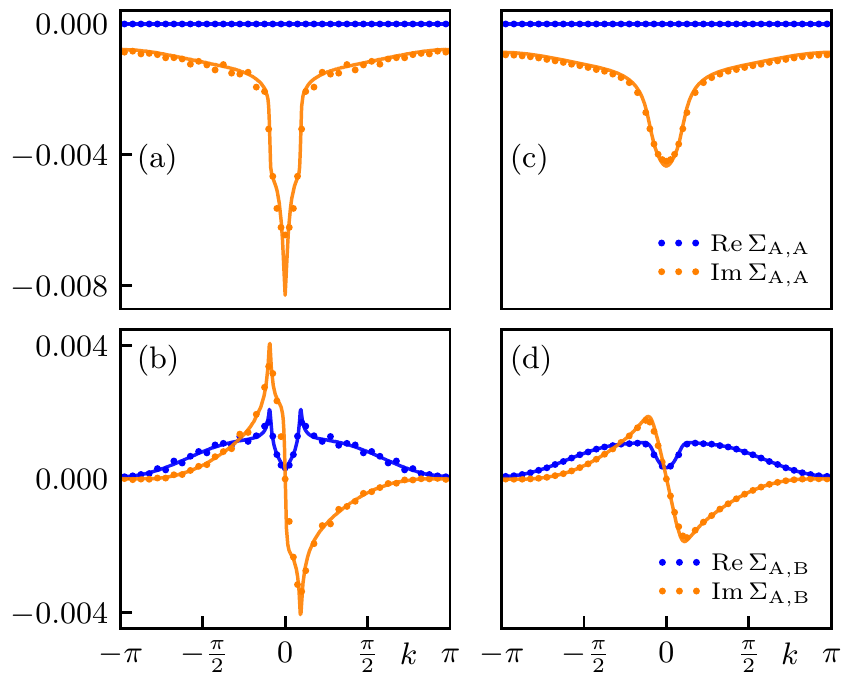}
\caption{Comparison of results for the self-energy at $\omega=0+\i \eta$ from analytic \gls{pt} (solid lines) and \gls{csba} (dots), for $U_\A = U_\B = 0.1$ and $\beta = 1$.
The value of $\eta$ corresponds to the regularization of the integral in \autoref{eqn:app:Sigma_ret}.
Panels (a) and (b) show $\Sigma_{\A,\A}$ and $\Sigma_{\A,\B}$, with $\eta=0.01$ for the fermionic \gls{pt} and $\eta=0$ for \gls{csba}.
Panels (c) and (d) instead use $\eta=0.1$ for both methods.
}
\label{fig:PT_vs_CSBA}
\end{figure}

\autoref{fig:PT_vs_CSBA} shows data for the self-energy components $\Sigma_{\A,\A}$ and $\Sigma_{\A,\B}$ ($\Sigma_{\B,\A}$ and $\Sigma_{\B,\B}$ follow from symmetry) obtained from \gls{csba} and fermionic plain \gls{pt} at $\omega=0$ for $U_\A = U_\B = 0.1$ and $\beta = 1$.
In panels (a) and (b), the regularization parameter for fermionic plain \gls{pt} is set to $\eta = 0.01$.
The quantitative discrepancies in the vicinity of $k=0$ are presumably an artefact from the regularization:
As illustrated in panels (c) and (d), evaluating the \gls{csba} result at $\omega=0+\i\eta$ and using a larger value of $\eta=0.1$ for both methods significantly improves the agreement.
The rationale behind this is that any discrepancies resulting from faulty damping of the \gls{csba} result at large $t$ should be overridden by the comparatively large value of $\eta$.

As long as $U_\A$, $U_\B$ are kept small and $\beta \lesssim 1$, the sample data shown here is representative of the agreement between \autoref{eqn:app:Sigma_ret} and the \gls{csba} result, meaning that there is always some deviation around $k = 0$ that can be mitigated by a finite $\eta$.

\section{Symmetries of \glspl{gf} at general complex argument}\label{sec:app:symmetries_kl}
The \nameKallen--Lehmann representation of the \gls{gf} can be written in terms of eigenvectors $\klc*{\Ket{E_{\mu}}}$ of the full Hamiltonian $H$,
\begin{align}
G_{j,l}(z) = \sum_{\mu,\nu} f_{\mu,\nu}(z) \brc{E_\mu }{c_j}{E_\nu }\brc{E_\nu }{c_l^\dagger}{E_\mu }, \label{eqn:app:kaellen_lehmann}
\end{align}
where the weights $f_{\mu,\nu}(z) = \frac{1}{Z} \frac{\powe{-\beta E_\mu} + \powe{-\beta E_\nu }}{z + E_\mu - E_\nu}$ have the symmetry $f_{\nu,\mu}(z) = -f_{\mu,\nu}(-z)$.
In general, complex conjugation $\K$ leads to 
\begin{align*}
G_{j,l}(z)^\ast &= \sum_{\mu,\nu} f_{\mu,\nu}(z^\ast) \brc{E_\nu }{c_j^\dagger}{E_\mu }\brc{E_\mu }{c_l}{E_\nu } = G_{l,j}(z^\ast),
\end{align*}
which in matrix notation reads $G(z)^\ast = G(z^\ast)^\transpose$, or
\begin{align}
G(z)^\dagger = G(z^\ast). \label{eqn:app:G_dagger_G_star}
\end{align}

\subsection{Time-Reversal Symmetry}\label{sec:app:trs}
Under \gls{trs}, fermionic operators transform as
\begin{align*}
\T c_j\T^{-1} &= \kla*{U_T}_{jj'}c_{j'} \quad \text{and} \quad \T c_j^\dagger\T^{-1} = \kla*{U_T^\ast}_{jj'}c_{j'}^\dagger,
\end{align*}
where repeated indices are implicitly summed over \cite{NHT_Classification_Chiu_2016}.
If the Hamiltonian is invariant under time-reversal, i.e. $\T H \T^{-1} = H$, the application of $\T$ to an eigenstate yields a (possibly different) eigenstate with the same energy: $\T \Ket{E_\nu} =: \Ket{\bar E_\nu}$.
Since $\T$ is anti-unitary, we have $\T=\U_T\K$ where $\U_T$ denotes the unitary part.
Anti-unitarity by definition implies $\brb{\T x}{\T y}  = \brb{x}{y}^\ast$ and thus $\brb{x}{\T^{-1} y}  = \brb{\T^{-1} \T x}{\T^{-1} y} = \brb{\T x}{y}^\ast$. For any two eigenstates of a \gls{trs}-invariant Hamiltonian, it follows that $\brc{E_\mu }{\T^{-1} A \T}{E_\nu }= \brc{\T E_\mu }{A}{ \T E_\nu }^\ast = \brc{\bar E_\mu }{A}{ \bar E_\nu }^\ast$ for any operator $A$. 
If the Hamiltonian is invariant under time-reversal, we can then rewrite \autoref{eqn:app:kaellen_lehmann} as
\allowdisplaybreaks
\begin{align*}
G_{j,l}(z) &= f_{\mu,\nu}(z)\brc{E_\mu }{\T^{-1} \T c_j \T^{-1} \T }{E_\nu }\\
&\hspace{1.53cm}\times\brc{E_\nu }{\T^{-1} \T  c_l^\dagger \T^{-1} \T }{E_\mu }\\
&= f_{\mu,\nu}(z) \brc{E_\mu }{\T^{-1} \kla*{U_T}_{j,j'} c_{j'} \T }{E_\nu }\\
&\hspace{1.53cm}\times \brc{E_\nu }{\T^{-1} \kla*{U_T^\ast}_{l,l'} c_{l'}^\dagger \T }{E_\mu }\\
&= \kla*{U_T^\ast}_{j,j'} \kla*{U_T}_{l,l'} f_{\mu,\nu}(z)\\
&\hspace{1.53cm}\times \brc{\bar E_\mu }{c_{j'}}{\bar E_\nu }^\ast \brc{\bar E_\nu }{c_{l'}^\dagger }{\bar E_\mu }^\ast\\
&= \kla*{U_T^\ast}_{j,j'} \kla*{U_T}_{l,l'} G_{j',l'}(z^\ast)^\ast,
\end{align*}
which in matrix notation reads $G(z) = U_T^\ast G(z^\ast)^\ast U_T^\transpose$, or equivalently $G(z)^\transpose = U_T G(z) U_T^\dagger$.

\subsection{Particle-Hole Symmetry}\label{sec:app:phs}
\gls{phs} is a unitary symmetry under which fermionic operators transform as \cite{NHT_Classification_Chiu_2016}
\begin{align*}
\C c_j\C^{-1} &= \kla*{U_C^\ast}_{jj'}c_{j'}^\dagger \quad \text{and} \quad \C c_j^\dagger\C^{-1} = \kla*{U_C}_{jj'}c_{j'}.
\end{align*}
Assuming invariance of the Hamiltonian under \gls{phs} implies that $\C \Ket{E_\nu} =: \Ket{\bar E_\nu}$ is an eigenstate with the same energy, and the \gls{gf} becomes
\begin{align*}
G_{j,l}(z) &= f_{\mu,\nu}(z) \brc{E_\mu }{\C^{-1} \C c_j \C^{-1} \C }{E_\nu }\\
&\hspace{1.53cm}\times \brc{E_\nu }{\C^{-1} \C c_l^\dagger \C^{-1} \C }{E_\mu }\\
&= f_{\mu,\nu}(z) \brc{E_\mu }{\C^{-1} \kla*{U_C^\ast}_{j,j'} c_{j'}^\dagger \C }{E_\nu }\\
&\hspace{1.53cm}\times\brc{E_\nu }{\C^{-1} \kla*{U_C}_{l,l'} c_{l'} \C }{E_\mu } \\
&= \kla*{U_C^\ast}_{j,j'} \kla*{U_C}_{l,l'} \kla*{-f_{\nu,\mu}(-z)}\\
&\hspace{1.53cm}\times \brc{\bar E_\mu }{c_{j'}^\dagger}{\bar E_\nu } \brc{\bar E_\nu }{c_{l'} }{\bar E_\mu }\\
&= -\kla*{U_C^\ast}_{j,j'} \kla*{U_C}_{l,l'} G_{l',j'}(-z),
\end{align*}
which can be written as $G(z) = -U_C G(-z) U_C^\dagger$.

\subsection{Chiral Symmetry}\label{sec:app:cs}
Under \gls{cs}, the combination of \gls{trs} and \gls{phs}, fermionic operators transform as \cite{NHT_Classification_Chiu_2016}
\begin{align*}
\S c_j\S^{-1} &= \kla*{U^\ast_S}_{jj'}c_{j'}^\dagger \quad \text{and} \quad \S c_j^\dagger \S^{-1} = \kla*{U_S}_{jj'}c_{j'}.
\end{align*}
Since $\S = \T \C$ is also anti-unitary, we once again have $\S=\U_S \K$ with some unitary part $\U_S$.
The unitary transformation matrix is related to the ones representing \gls{trs} and \gls{phs} via $U_S = U_C^\ast U_T$.
The \gls{gf} thus satisfies
\begin{align*}
G_{j,l}(z) &= f_{\mu,\nu}(z) \brc{E_\mu }{\S^{-1} \S c_j \S^{-1} \S }{E_\nu }\\
&\hspace{1.53cm}\times\brc{E_\nu }{\S^{-1} \S c_l^\dagger \S^{-1} \S }{E_\mu }\\
&= f_{\mu,\nu}(z)\brc{E_\mu }{\S^{-1} \kla*{U_S^\ast}_{j,j'} c_{j'}^\dagger \S }{E_\nu }\\
&\hspace{1.53cm}\times\brc{E_\nu }{\S^{-1} \kla*{U_S}_{l,l'} c_{l'} \S }{E_\mu }\\
&= \kla*{U_S}_{j,j'} \kla*{U_S^\ast}_{l,l'} \kla*{-f_{\nu,\mu}(-z)}\\
&\hspace{1.53cm}\times \brc{\bar E_\mu }{c_{j'}^\dagger}{\bar E_\nu }^\ast \brc{\bar E_\nu }{c_{l'} }{\bar E_\mu }^\ast \\
&= -\kla*{U_S}_{j,j'} \kla*{U_S^\ast}_{l,l'} G_{l',j'}(-z^\ast)^\ast,
\end{align*}
which is equivalent to $G(z) = -U_S G(-z^\ast)^\dagger$ or, via \autoref{eqn:app:G_dagger_G_star}, $U_S^\dagger = -U_S G(-z) U_S^\dagger$.

\subsection{Summary}
To summarize, assuming the respective symmetry, the \gls{gf} for general complex $z$ satisfies
\begin{align}
\text{\gls{trs}} \quad &: & G(z)^\transpose &= U_TG(z) U_T^\dagger, \\
\text{\gls{phs}} \quad &: & G(z)^\transpose &= -U_CG(-z) U_C^\dagger,\\
\text{\gls{cs}} \quad &: & G(z) &= -U_SG(-z) U_S^\dagger
\end{align}
or any equivalent forms obtained by inserting the symmetry-independent result from \autoref{eqn:app:G_dagger_G_star} (i.e. $G(z)^\dagger = G(z^\ast)$).
At zero temperature, the same symmetry constraints have already been derived by \cite{GF_Symmetry_Gurarie_2011}. 
The translation to the Matsubara \gls{gf} is trivial by replacing $z=\i\omega_n^\F$.
For the real time \gls{gf}, \autoref{eqn:app:G_dagger_G_star} yields $G^\ret(\omega)^\dagger = G^\text{adv}(\omega)$.
Then, \gls{cs} for the retarded \gls{gf} can be obtained by $z=\omega+\i\eta$, which gives $G^\ret(\omega) = -U_SG^\text{adv}(-\omega) U_S^\dagger = -U_S G^\ret(-\omega)^\dagger U_S^\dagger$.

\bibliography{bibliography}

\end{document}